%% file: twhs_nf22.tex
\documentclass[10pt]{iopart}
\usepackage[ngerman,english]{babel}
\usepackage{graphicx}
\usepackage{subcaption}

\expandafter\let\csname equation*\endcsname\relax
\expandafter\let\csname endequation*\endcsname\relax
\usepackage{amsmath,amssymb}
\usepackage{bm}
\usepackage{siunitx}
\usepackage{isotope}
\usepackage{mhchem}
\usepackage{letltxmacro}
\makeatletter
\let\oldr@@t\r@@t
\def\r@@t#1#2{%
  \setbox0=\hbox{$\oldr@@t#1{#2\,}$}\dimen0=\ht0
  \advance\dimen0-0.2\ht0
  \setbox2=\hbox{\vrule height\ht0 depth -\dimen0}%
  {\box0\lower0.4pt\box2}}
\LetLtxMacro{\oldsqrt}{\sqrt}
\renewcommand*{\sqrt}[2][\ ]{\oldsqrt[#1]{#2}}
\makeatother

\usepackage{mathtools}

\DeclarePairedDelimiter\round{\lfloor}{\rceil}
\newcommand{\IPP}{Max-Planck-Institut f\"ur Plasmaphysik, Boltzmannstr.~2, 85748 Garching, Germany}
\newcommand{\IPG}{Max-Planck-Institut f\"ur Plasmaphysik, Wendelsteinstr.~1, 17491 Greifswald, Germany}

\input{commands}
\graphicspath{{images/}}

\begin{document}
\title{Multi-scale analysis of global electromagnetic instabilities in ITER Pre-Fusion-Power Operation plasmas}
\author{T.~Hayward-Schneider$^1$,
Ph.~Lauber$^1$,
A.~Bottino$^1$,
A.~Mishchenko$^2$}
\address{$^1$ \IPP\\$^2$ \IPG}
\ead{thomas.hayward@ipp.mpg.de}
%\submitto{\NF}
  \input{abstract}
  \maketitle
  \ioptwocol % for 2-col
%  \pagenumbering{arabic}

%\begin{multicols}{2}
  \input{introduction}

  \input{model}
  \input{setup}
  \input{results}
%  \input{results-nl}
  \input{conclusions}
  \input{acknowledgements}
  \input{appendix}
  \section*{References}
  \bibliographystyle{unsrt}
  \bibliography{twhs_nf22}
%\end{multicols}{2}
\end{document}

%% file: commands.tex
\newcount\colveccount
\newcommand*\colvec[1]{
        \global\colveccount#1
        \begin{pmatrix}
        \colvecnext
}
\def\colvecnext#1{
        #1
        \global\advance\colveccount-1
        \ifnum\colveccount>0
                \\
                \expandafter\colvecnext
        \else
                \end{pmatrix}
        \fi
}

\newcommand{\grad}[2][]{\nabla_{#1} #2}

\newcommand{\dif}{\mathop{}\!\mathrm{d}}

\newcommand{\df}{\textrm{d}}

\newcommand{\mun}{\mu_\textnormal{0}}
\newcommand{\unitvecb}{\boldsymbol{b}}
\newcommand{\vpar}{v_{\parallel}}
\newcommand{\vpardot}{\dot{v}_{\parallel}}
\newcommand{\Apar}{A_{\parallel}}
\newcommand{\Aparh}{A_{\parallel}^{(h)}}
\newcommand{\Apars}{A_{\parallel}^{(s)}}
\newcommand{\bparstar}{B_{\parallel}^{*}}
\renewcommand{\vec}[1]{\boldsymbol{#1}}
\newcommand{\pder}[2][]{\frac{\partial#1}{\partial#2}}

\newcommand{\der}[2][]{\frac{\dif#1}{\dif#2}}

\newcommand{\Alfv}{Alfv\'en}

\newcommand{\omegaA}{\omega_\textnormal{A}}
\newcommand{\omegaci}{\omega_\textnormal{ci}}
\newcommand{\wci}{\omegaci}
\newcommand{\wa}{\omegaA}
\newcommand{\qmax}{q_\textrm{max}}

\newcommand{\trm}[1]{\textrm{#1}}

%% file: abstract.tex
Global electromagnetic gyrokinetic simulations are performed with the Particle-in-Cell code ORB5 for an ITER Pre-Fusion-Power-Operation (PFPO) plasma scenario, with half-field (\SI{2.65}{\tesla}) and half-current (\SI{7.5}{\mega \ampere}).
We report on a `multi-scale` analysis of the discharge, considering eigenmodes and instabilies across three scale-lengths.
Although the scenario will nominally have neutral beam heating with particles injected with \SI{1}{\mega \electronvolt}, \Alfv{} eigenmodes are investigated in the absence of such source, and Reversed Shear (RSAE), Toroidal (TAE) and Elliptical (EAE) \Alfv{} eigenmodes are found with weak damping for moderately low toroidal mode numbers ($10 \leq n \leq 35$).
At higher toroidal mode numbers ($40 \leq n \leq 70$), unstable \Alfv{}ic modes have been observed close to rational surfaces and are labelled as Beta-induced \Alfv{} eigenmodes (BAE)/\Alfv{}ic Ion Temperature Gradient (AITG) modes, since their frequency is associated with the BAE gap and they are driven by the bulk plasma on the \Alfv{}ic continuum.
These modes are unstable in the absence of energetic particles, and adding a species of energetic particles (with an isotropic \SI{1}{\mega \electronvolt} slowing down distribution) has negligible impact on their growth rate.
At higher toroidal mode numbers ($150 \lessapprox n < 220$), low frequency microscale instabilities are observed.

%% file: introduction.tex
\section{Introduction}
Energetic particles (EPs), such as Neutral Beam Injection (NBI) or alpha particles generated by fusion reactions, can have a significant impact on the stability of electromagnetic instabilities in tokamak plasmas.
In turn, these instabilities can lead to energetic particle transport, or losses, affecting the efficiency of the plasma heating, or potentially increasing wall heat loads.
In addition, the bulk plasma can also interact with these electromagnetic instabilities, not only contributing to the damping, but in some cases also driving these instabilities.

In previous work, we studied the effect of alpha particle drive on \Alfv{} eigenmodes in a $Q=10$ scenario for ITER.
In this paper, we consider a broad range of electromagnetic instabilities potentially present in the pre-fusion-power-operation (PFPO) phase of ITER.
Specifically, we consider a scenario with a half-field, half-current, hydrogen plasma, and \SI{1}{\mega \electronvolt} neutral beam injection (NBI).
We consider the \Alfv{} eigenmodes which may potentially be destabilized by the NBI ions, and also smaller scale instabilities, namely meso-scale \Alfv{}ic modes such as Beta-induced \Alfv{} eigenmodes/\Alfv{}ic Ion Temperature Gradient modes (BAE/AITG), as well as microscale instabilies such as those associated with (electromagnetic) turbulence.

In this work, we use the global electromagnetic gyrokinetic particle-in-cell code ORB5~\cite{Lanti2020, Mishchenko2019} for the simulations, which are also compared to \Alfv{} continuum calculations using the linear global gyrokinetic eigenvalue solver LIGKA~\cite{Ligka}.

This paper is organized as follows:
\S\ref{sec:model} describes the theoretical model used for the calculations, and the numerical implementation of the numerical tool ORB5;
\S\ref{sec:iter} outlines the scenario being studied in this work;
\S\ref{sec:results} presents the results, split into three sections: linear \Alfv{} eigenmodes which might be driven unstable by EPs, unstable AEs driven by the bulk plasma, and instabilities associated with microturbulence. Finally, \S\ref{sec:conc} presents a summary and outlook of the paper.

%% file: model.tex
\section{Physical model}
\label{sec:model}
In this work, the numerical results presented are obtained using the ORB5 code~\cite{Lanti2020}. This is a global electromagnetic gyrokinetic particle-in-cell (PIC) code which uses markers to sample the 5D phase space ($\vec{R}$, $v_\parallel$, $\mu$), with the equations independent of the gyrophase, and the magnetic moment ($\mu$) of a marker constant in the absence of collisions.
%\begin{equation}
%        \delta f_s(\vec{R},v_\parallel,\mu,t)
%        =
%        \sum_{i=1}^{N_p} w_{s,i}(t)\delta(\vec{R} - \vec{R}_i) \delta(v_\parallel - v_{\parallel,i}) \delta(\mu - \mu_i)
%\end{equation}

The perturbed part of the distribution function is then evolved according to the gyrokinetic Vlasov equation
\begin{equation}
        \der{t}{\delta f_s}
        =
        -\dot{\vec{R}} \cdot \left.\frac{\partial F_{0s}}{\partial \vec{R}}\right|_{\epsilon,\mu} -
        \dot{\epsilon} \cdot \left.\frac{\partial F_{0s}}{\partial \epsilon}\right|_{\vec{R},\mu}
\end{equation}
(where subscript $s$ denotes a plasma species $s$) according to the equations for the particle equations of motion
\begin{multline}
\label{eq:Rdot}
        \dot{\vec{R}} = \vpar \unitvecb - \vpar^2\frac{c m_\trm{s}}{q\bparstar}\vec{G} + \mu \frac{Bcm_\trm{s}}{q\bparstar}\unitvecb \times \frac{\grad{B}}{B} + \\
        \frac{\unitvecb}{\bparstar} \times \nabla \big\langle \phi - \vpar \Apar^{\trm{s}} - \vpar \Apar^{\trm{h}} \big\rangle - \frac{q_\trm{s}}{m_\trm{s}} \langle \Apar^{\trm{h}} \rangle \unitvecb^{*}
\end{multline}
\begin{multline}
\label{eq:vdot}
        \vpardot = \mu B \nabla \cdot \unitvecb + \mu \vpar \frac{c m_\trm{s}}{q_\trm{s} \bparstar}\vec{G}\cdot\grad{B} \
        - \mu \frac{\unitvecb \times \grad{B}}{\bparstar} \cdot \grad{\langle \Apar^{(s)} \rangle} \\
        -\frac{q_\trm{s}}{m_\trm{s}}\left[\unitvecb^{*}\cdot\nabla \big\langle \phi - \vpar \Apar^{(h)} \big\rangle + \pder{t}{}\langle\Apar^{(s)} \rangle \right]
\end{multline}
\begin{equation}
\label{eq:epsdot}
        \dot{\epsilon} = \vpar \vpardot + \mu \grad{B} \cdot \dot{\vec{R}}
\end{equation}

where
        $\vec{B} = \nabla \times \vec{A}$, $\unitvecb = \vec{B} / B$, $\bparstar = \unitvecb \times \vec{A}^{*}$
\begin{equation*}
        \vec{A}^{*} = \vec{A} + \left( \frac{m_\trm{s} c}{q_\trm{s}} \vpar + \langle \Apar^{(s)} \rangle \right) \unitvecb
\end{equation*}
\begin{equation*}
        \unitvecb^{*} = \frac{\nabla \times \vec{A}^{*}}{\bparstar} = \unitvecb - \left( \frac{c m_\trm{s}}{q \bparstar} \vpar + \langle \Apar^{(s)} \rangle \right) \vec{G}
\end{equation*}
\begin{equation*}
        \vec{G} = \unitvecb \times (\unitvecb \times (\nabla \times \unitvecb))
\end{equation*}
and with the gyroaveraged potential $\langle \phi \rangle = \oint \phi(\vec{R} + \vec{\rho})\df \alpha/(2\pi)$ with $\rho$ the gyroradius of the particle and $\alpha$ the gyro-phase.

These equations are coupled to the field equations, the linearized gyrokinetic quasineutrality equation
\begin{equation}
        - \nabla \cdot \left[ \left( \sum_{s} \frac{q_\trm{s}^2 n_\trm{s}c^2}{T_\trm{s}} \rho_\trm{s}^2\right) \nabla_{\perp} \phi \right] = \sum_{s} q_\trm{s} n_\trm{1s}
\end{equation}
For the parallel Amp\`ere's law, the perturbed magnetic potential $\Apar$ is split in to the symplectic part $\Apar^{(s)}$, which is found from
\begin{equation}
        \label{eq:Ohm}
        \pder{t} \Apar^{(s)} + \unitvecb \cdot \grad{\phi} = 0
\end{equation}
and the Hamiltonian part $\Apar^{(h)}$, solved using the mixed-variable parallel Amp\`ere's law,
\begin{equation}
        \left(\sum_{s} \frac{\beta_\trm{s}}{\rho_\trm{s}^2} - \nabla_{\perp}^2 \right) \Aparh = \mun \sum_{s} j_{\parallel 1 s} + \nabla_{\perp}^2 \Apars
\end{equation}
with $n_\trm{1s} = \int \df{}^6 Z \delta f_\trm{s} \delta(\vec{R}+ \vec{\rho} - \vec{x})$ the perturbed gyrocenter density, $\rho_\trm{s} = \sqrt{m_\trm{s} T_\trm{s}}/(q_\trm{s} B)$ the thermal gyroradius, $q_\trm{s}$ the particle charge, $\df{}^6Z = \bparstar \df{} \vec{R}\df{} \vpar \df{} \mu \df{} \alpha$ the phase space volume, $j_\trm{1s} = \int \df{}^6 Z \vpar \delta f_\trm{s} \delta(\vec{R} + \vec{\rho} - \vec{x})$ the perturbed parallel gyrocenter current. Note that due to the scale separation between ion and electron Larmor radii, electrons are treated differently from ion species, and terms with $\rho_\trm{e}$ are neglected.

The particular method of using the mixed-variables formulation, and then applying a pullback procedure to the particles is described in reference~\cite{Mishchenko2019}, where its advantages for solving electromagnetic equations are discussed.

A set of straight field line coordinates are used, with radial coordinate $s=\sqrt{\psi/\psi_\trm{edge}}$ (where $\psi$ is the poloidal flux), toroidal angle $\varphi$, and poloidal angle
\[\theta^* = \frac{1}{q(s)} \int_0^\theta \frac{\vec{B}\cdot \grad\varphi}{\vec{B}\cdot \grad {\theta^\prime}} \df \theta^\prime\]
(where $\theta$ is the geometric poloidal angle, and q(s) the safety factor)\footnote{$\theta^*$ in ORB5 is often labelled elsewhere as $\chi$.}.

The field equations are solved on a basis of cubic finite elements and using Fourier methods in the two angular directions. This allows the use of Fourier filtering, which helps to reduce noise by excluding non-physical modes which are far from being field aligned~\cite{McMillan2010}. The local poloidal Fourier filter will retain modes with poloidal mode numbers $m_n(s) = \round{n q(s)} \pm \Delta m$ at each radial point for each toroidal mode n, where a typical values of $\Delta m$ is 5.

The distribution function $f(\vec{R}, \vpar, \mu, t)$ is separated into a time-independent background part ($F_0(\Psi=s^2, E=\frac{1}{2}\vpar^2 + \mu B, \vpar)$), and a time dependent part $\delta f(\vec{R}, \vpar, \mu, t)$. The time dependent part of the distribution function is represented by markers, using the so-called particle-in-cell (PIC) Lagrangian method, with each ``marker'' representing a small 5D volume of phase-space, and carrying a particle weight, $w(t)$.
These markers are evolved using the equations of motion for the gyrocenters, and with a weight equation to evolve the weights.
The fields are solved from the charge and current densities, calculated by depositing the charge/current from the markers using the gyrocenters for electron markers, and gyrorings (using in this work a 4-point average) for the ion markers.

The system of particles and fields is evolved using a 4th-order Runge-Kutta timestepping scheme.

Linear simulations can be performed by linearizing equations for the particle trajectories and weights, equations~\ref{eq:Rdot},~\ref{eq:vdot},~and~\ref{eq:epsdot}.

%% file: setup.tex
\section{Scenario description}
\label{sec:iter}
\begin{figure}
  \centering
  \includegraphics[width=0.39\textwidth]{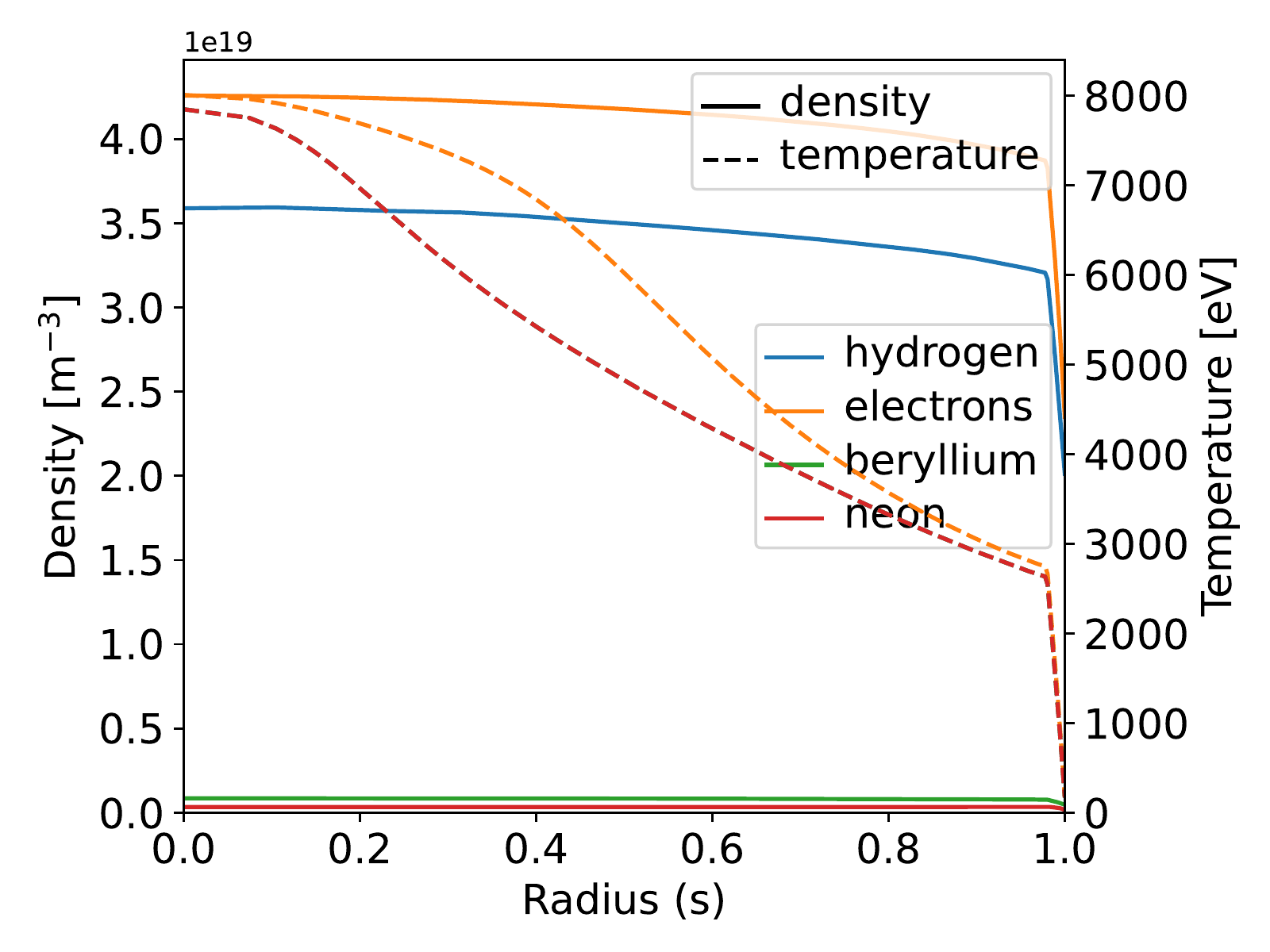}
  \includegraphics[width=0.45\textwidth]{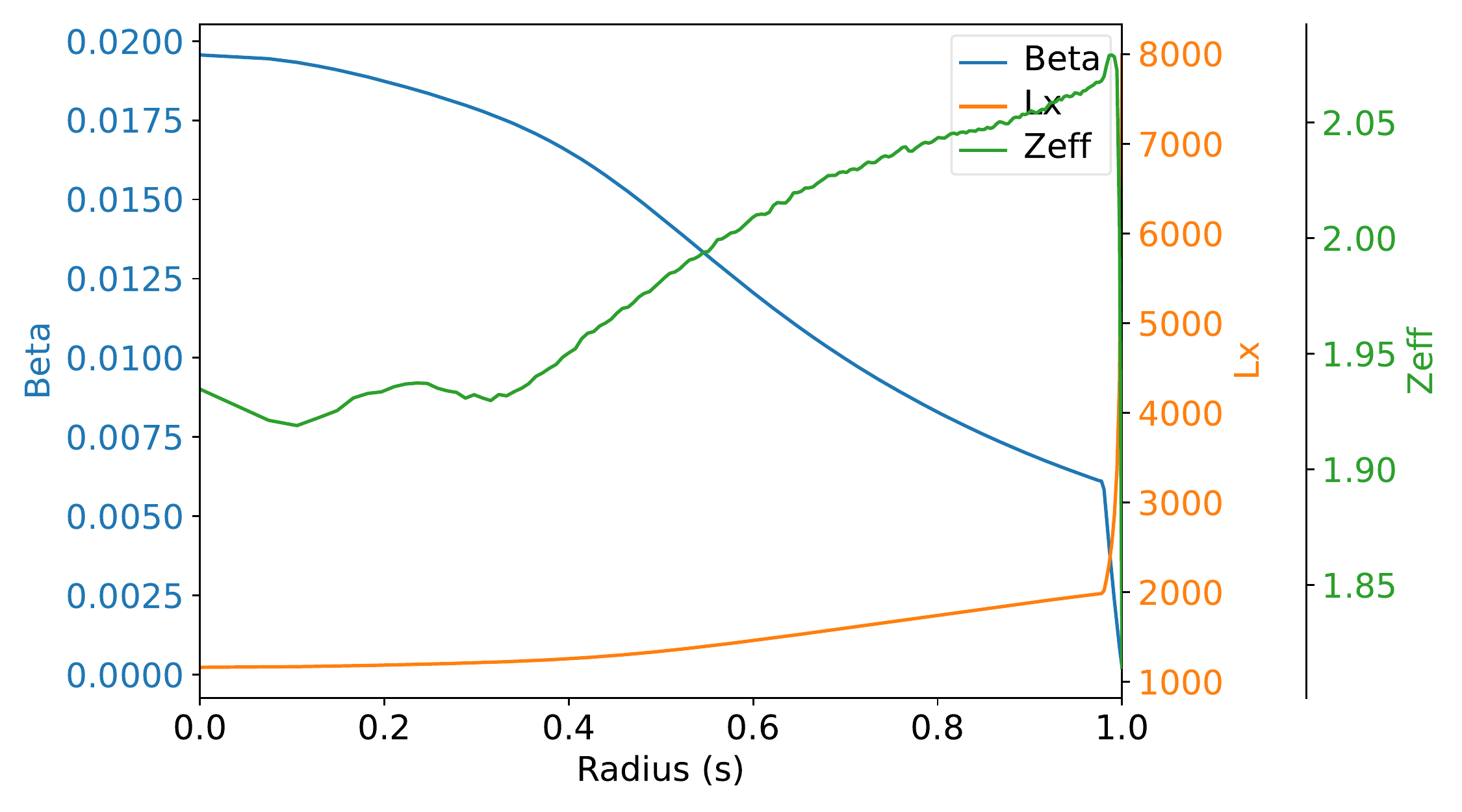}
  \caption{(a) Density (solid) and temperature (dashed) profiles for the bulk species in the scenario; \\
(b) profiles of dimensionless quantities: electron $\beta(s)$, $L_\textrm{x}(s) = 2/\rho^*(s) = 2 a/\rho_s(s)$, $Z_\textrm{eff}(s)$.}
  \label{fig:iter_eq2}
\end{figure}

We consider a half-field, half-current ITER PFPO~\cite{Polevoi2021, Loarte2021} scenario, modelled using ASTRA~\cite{ASTRA} and made available at ITER through the Integrated Modelling \& Analysis Suite (IMAS)~\cite{Imbeaux2015}.
The major radius, $R=\SI{6.2}{\m}$; the minor radius, $a=\SI{2.0}{\m}$; the magnetic field on axis $B_0=\SI{2.65}{\tesla}$; the bulk ions are \ce{H}, with some additional Beryllium and Neon impurities ($< 1\%$).
The electron temperature profile is peaked, with the electron temperature on axis $T_{\trm{e},s=0} \approx \SI{8}{\kilo \electronvolt}$, and the electron temperature at the top of the pedestal $(T_\trm{e,ped. top} \approx \SI{3}{\kilo \electronvolt}$).
The main ion temperature profile has very similar core and pedestal top values, but the peaking profile is slightly different, with the steeper gradient region closer to the core than in the case of the electron temperature.
Impurity ion species are taken to have the same temperature profiles as the main ion species.
The density profiles of the ions and electrons (linked via quasineutrality and the impurity density profiles) are almost flat, but marginally peaked.
The on-axis electron density $n_{\trm{e},s=0} \approx \SI{4e19}{m^{-3}}$.
The temperature and density profiles for the background species are shown in figure~\ref{fig:iter_eq2}(a).

These parameters give an \Alfv{} frequency ($\wa$) on axis of
\SI{1.58e6}{\radian \per \second}, approximately $\wci/177$.

\begin{figure}
  \centering
  \includegraphics[width=0.39\textwidth]{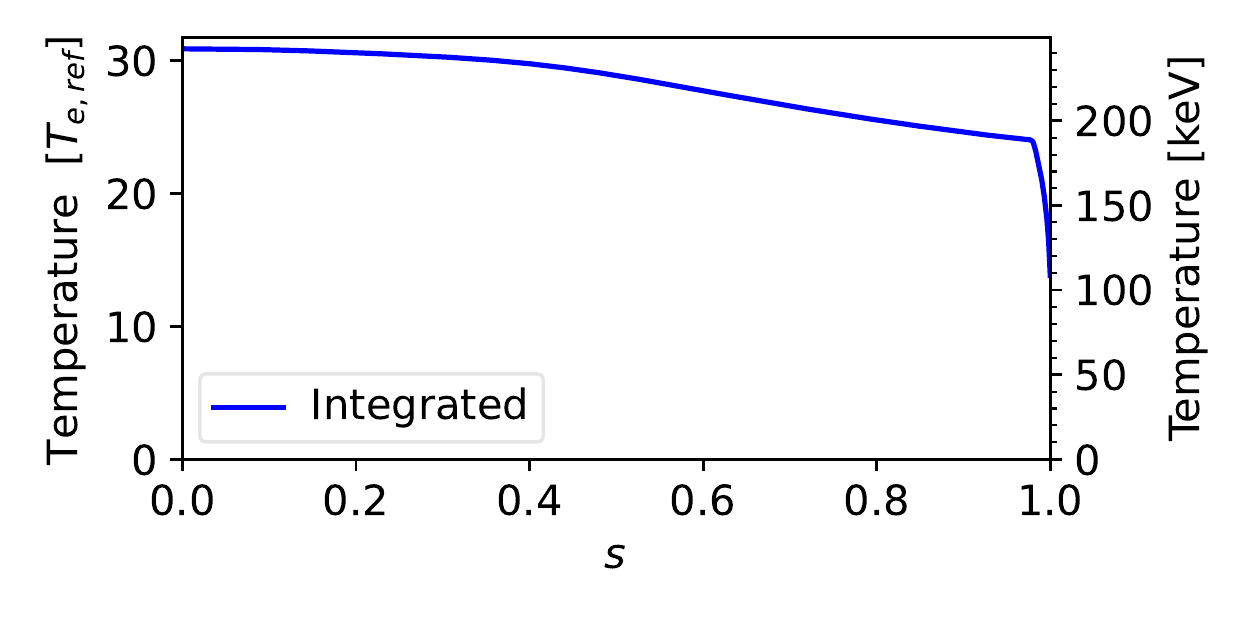}
  \caption{Energetic particle effective temperature profile, obtained by integrating the moments of the distribution function output from a simulation.}
  \label{fig:iter_T_nbi}
\end{figure}

The energetic particles in the scenario are from a negative-ion neutral beam injection (NBI, or N-NBI) source with a single birth energy of \SI{1}{\mega \electronvolt}.
In this work, when considered, the energetic particles, are modelled using an isotropic slowing down distribution~\cite{Vannini2022} with birth energy \SI{1}{\mega \electronvolt}.
If we calculate the equivalent temperature by integrating moments of this distribution function, we find that this corresponds to \SI{242}{\kilo \electronvolt} or $T_\textrm{EP}/T_\textrm{e} = 30.9$ on axis, with the ratio decreasing to about $24$ at the top of the pedestal.
The profile of the equivalent temperature for the NBI is shown in figure~\ref{fig:iter_T_nbi}.
While this neglects the anisotropy present in a realistic distribution function (also considered in the ASTRA model), we leave an anisotropic analytical distribution function~\cite{Rettino2022}, or a numerical treatment of the NBI from ASTRA's Fokker-Plank solver to future work.
In fact, we leave most of the modelling of drive from NBI ions to future work.
%In figure~\ref{fig:ep_profiles}, we show the density and temperature for a sim

%The energetic particles, a population of fusion alpha particles with birth energy $3.5\ \trm{MeV}$ are peaked in the core of the plasma where $n_{\trm{EP},s=0}/n_{\trm{e},s=0} \approx 0.75 \%$.
%The energetic particle density profile is shown in figure~\ref{fig:iter_eq}(c).

The profiles and equilibrium considered were taken from IMAS at ITER, the scenario in question being shot 101006, run 50.
The equilibrium used was generated by running the CHEASE MHD equilibrium code~\cite{CHEASE} reading from IMAS, and outputting an file using the CHEASE-ORB5 interface. \footnote{CHEASE takes the equilibrium from IMAS/IDS as an input, but converges to its own solution. As a result, the q-profile obtained in the output of CHEASE is not identical to the q-profile from the IDS, as written by ASTRA.}

\begin{figure}
  \centering
  \includegraphics[width=0.35\textwidth]{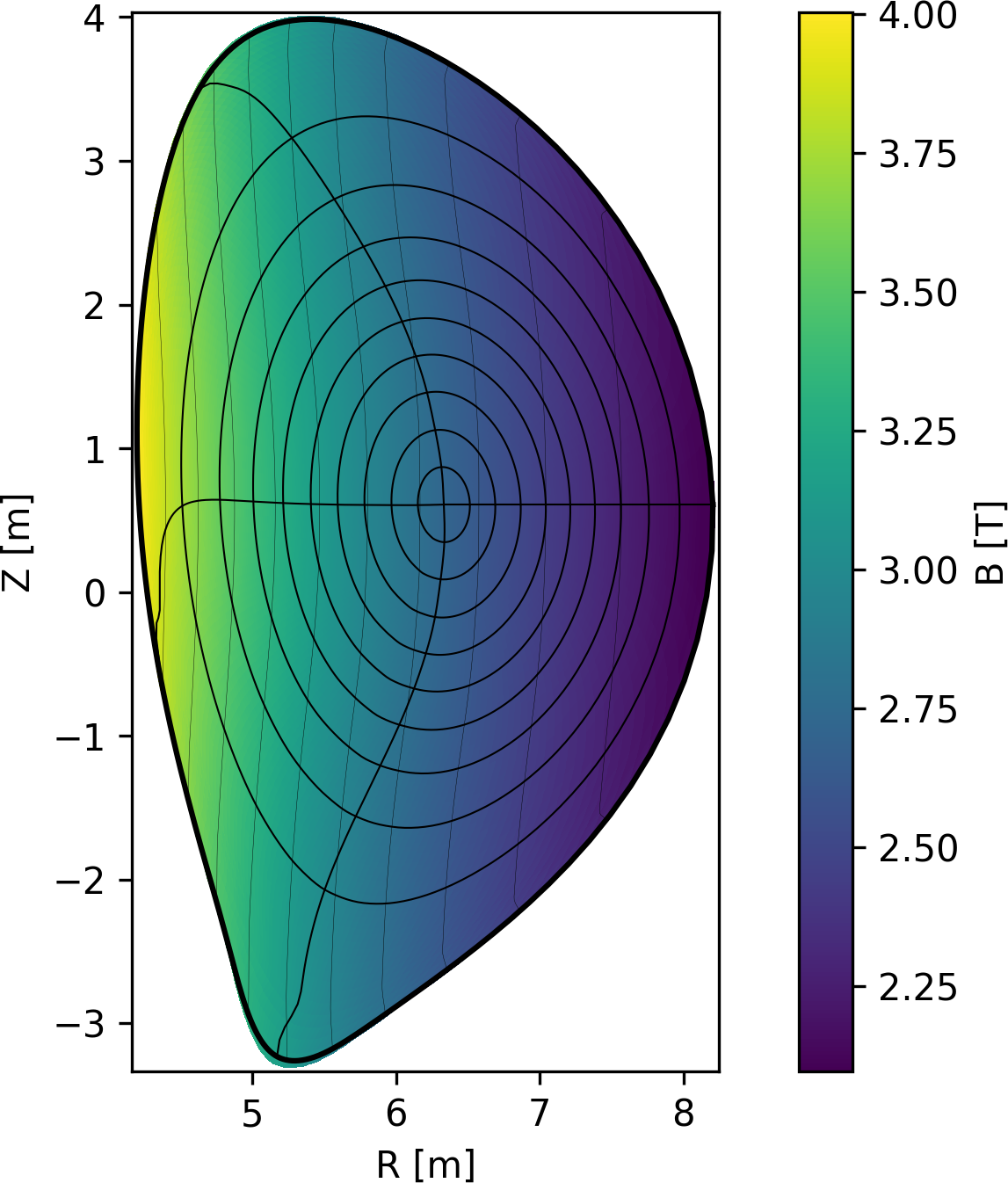}
  \includegraphics[width=0.4\textwidth]{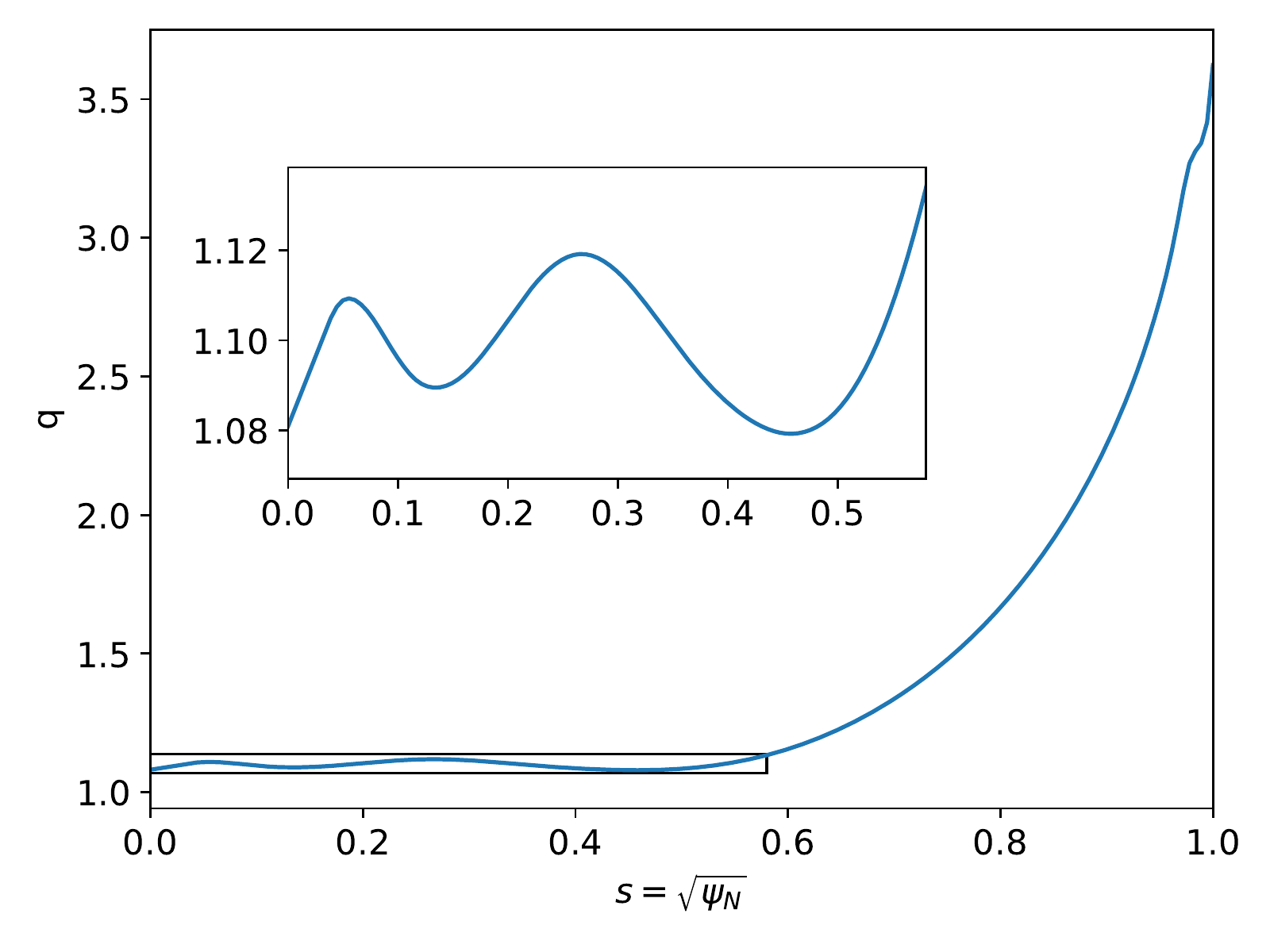}
  \caption{(a) ITER 15MA scenario equilibrium. The colour scale indicates $|B|$, the nested rings are surfaces of poloidal flux (equispaced in $s=\sqrt{\Psi_\trm{N}}$), the 4 lines which meet approximately at right angles in the centre are lines of constant straight-field-line poloidal angle ($\theta^*$), the thin almost vertical lines are equispaced contours of $|B|$; \\
(b) the safety factor profile (with inset zoom showing the values in the inner part of the tokamak).}
  \label{fig:iter_eq}
\end{figure}

Due to the steep gradients at the edge of the plasma, we consider at most the region of the plasma $0 \leq s \leq 0.9$, which corresponds to the outermost interior ring plotted in figure~\ref{fig:iter_eq}(a).

Finally, while all simulations presented here are electromagnetic and performed with kinetic electrons, we increase the electron mass in order to reduce the numerical cost.
In previous work~\cite{Hayward-Schneider2021}, it was found that the mode damping was only weakly affected by increasing the electron mass, but that the numerical savings allowed more confidence in the numerical timestep convergence.
Therefore, unless otherwise specified, the ratio of $m_\textrm{H}/m_\textrm{e}$ shall be taken as 400.

%% file: results.tex
\section{Results}
\label{sec:results}
By changing the toroidal mode number(s) in the Fourier filter, it is possible to perform simulations of reduced systems, and thereby separate the physics scales of interest: of particular interest for linear studies.
Unless stated otherwise, the simulations presented retain only a single toroidal mode number in the filter, and are initialized with a density perturbation (a perturbation on the particle weights depending only on the spatial coordinates) corresponding to the same toroidal mode number.
The radial and poloidal structure of the perturbation may vary, but is typically either radially broad and approximately field aligned (as is used in this work), or radially narrower, with e.g.\ a pair of poloidal harmonics designed to initialize the TAE, EAE, etc, in which case it will be radially located at the corresponding rational/half-rational surface of the TAE/EAE.

\subsection{\Alfv{} eigenmodes}
\label{sec:results_aes}
We start by considering the range of \Alfv{} eigenmodes which might reasonably be expected to be driven unstable by the energetic particles, focussing on $10 < n < 35$.

First, we show $n=12$, for which we perform a simulation with reduced mass ratio $m_\textrm{H}/m_\textrm{e}=400$.

Running a simulation with $N_p = (32,128,8,8)\times 10^6$ markers for the hydrogen, electrons, beryllium, and neon species respectively. The simulation is performed with $\Delta t =$\SI{10}{\wci^{-1}}, and $\Delta m = 5$, with a spatial grid resolution of $(N_s, N_{\theta^*}, N_\varphi) = (1024, 192, 64)$.

\begin{figure}
  \centering
  \includegraphics[width=0.25\textwidth]{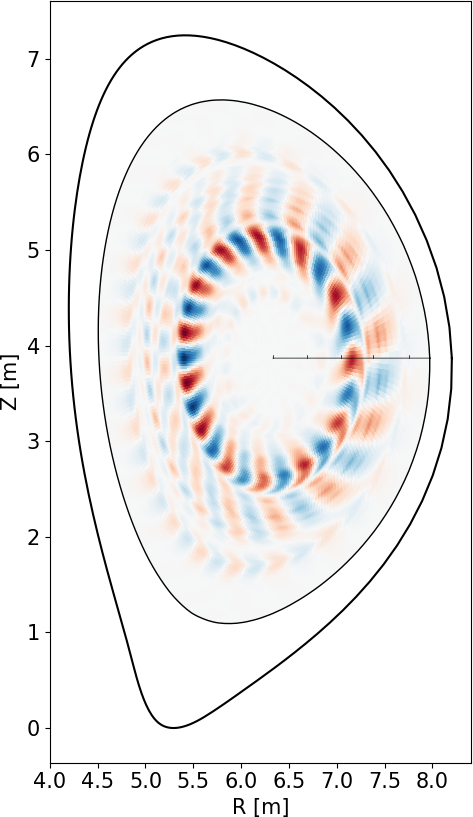}
  \caption{Poloidal cross section of the electrostatic potential for a simulation with $n=12$.}
  \label{fig:n12_psc}
\end{figure}

%We take the opportunity to introduce some diagnostics.
In figure~\ref{fig:n12_psc}, we show the poloidal cross section of the electrostatic potential at $t=$\SI{40000}{\wci^{-1}}.
%Considering the time evolution, we plot the evolution of $||\Phi_{n,m}}(r)||_\infty(t)$ for the poloidal harmonics in the simulation (those seen in figure~\ref{fig:n12_m_of_s}).

\begin{figure}
  \centering
  \includegraphics[width=0.4\textwidth]{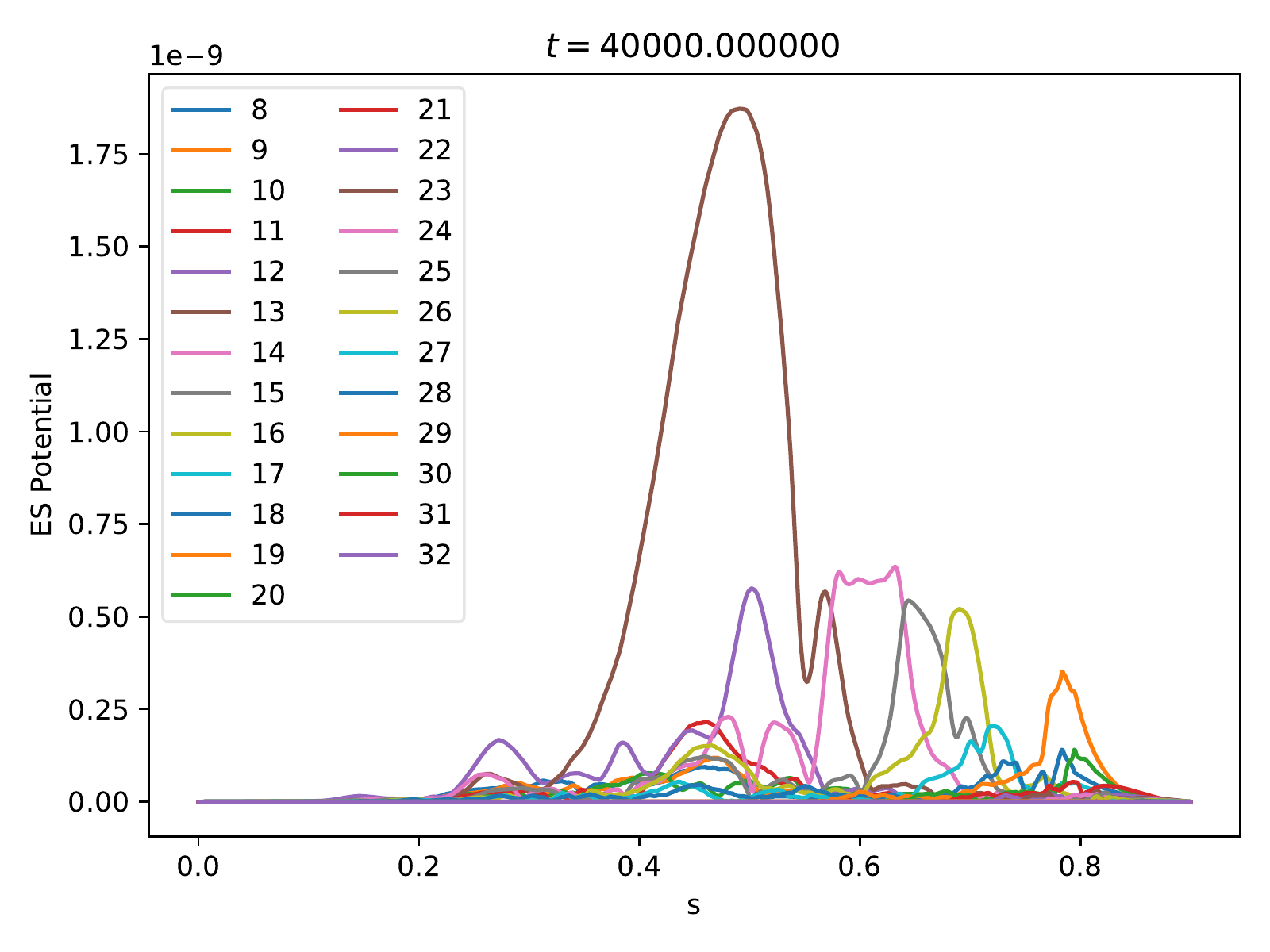}
  \caption{Plots of the absolute values of radial profiles of the $(n,m)=(12,m)$ helicities of the electrostatic potential, with the legend corresponding to $m$ (approximately increasing from core to edge with $q(s)$).}
  \label{fig:n12_m_of_s}
\end{figure}

\begin{figure}
  \centering
  \includegraphics[width=0.45\textwidth]{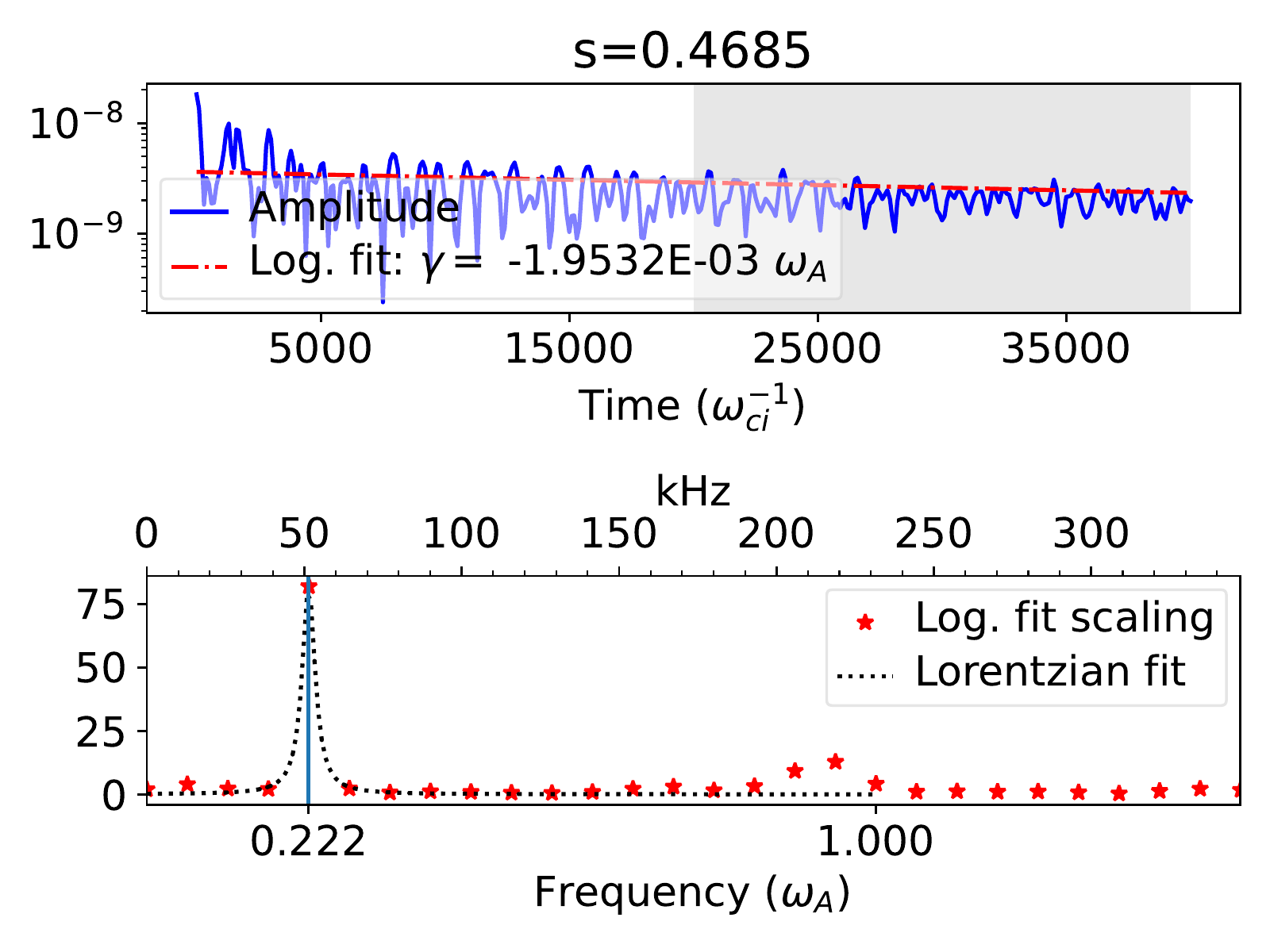}
  \caption{ For the point $s=0.4685$: (a) Plot of the $n=12$ envelope, with an exponential fit to the extrema in the range $20000 < t \wci < 40000$ (shaded grey). (b) Discrete Fourier transform of the electrostatic potential after dividing by the fitted exponential in (a).}
  \label{fig:n12_freq_growth}
\end{figure}

For the frequency evolution, we show in figure~\ref{fig:n12_freq_growth} how we fit the frequency, taking the example of $s=0.4685$ (the location of the peak in figure~\ref{fig:n12_m_of_s}).
In the upper panel, we plot the evolution of the envelope of the toroidal mode $\left|\Phi_{n,s}(t)\right|$.
Fitting this from $t>$\SI{20e3}{\wci^{-1}}, we obtain a growth rate of $\gamma=$\SI{-1.95e-3}{\wa}.
We can now normalize the signal by multiplying $\Re(\Phi_{n,s}(t))$ by $e^{-\gamma t}$~\footnote{Taking the real part here is optional when dealing with a complex signal such as the toroidal/poloidal Fourier coefficients of the field. In such a case, we can refrain from taking the real part, and then the resulting analysis will allow us to differentiate between clockwise and anti-clockwise rotating modes. In cases where we analyse a real signal, however, the positive and negative frequency components in the output will necessarily be complex conjugates of each other, and the sign therefore has no meaning.}.
We then perform a Discrete Fourier transform of this normalized signal (again, for $t>$\SI{20e3}{\wci^{-1}}), the results of which are in the lower panel. In this case, we find ($\omega$, $\gamma$) = $(0.222, -1.95 \times 10^{-3})$ $\omega_A$, or $f =$ \SI{51.4}{\kilo \hertz} ($\gamma/\omega =$~\SI{-0.88}{\%}).
Such weak damping is sufficiently small that we might possibly expect these modes to be driven unstable when energetic particles are added~\cite{Hayward-Schneider2021}.

\begin{figure}
  \centering
  \includegraphics[width=0.45\textwidth]{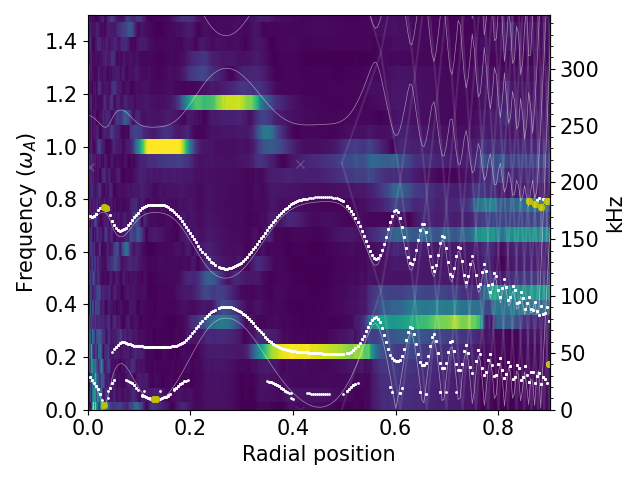}
  \caption{Radial frequency spectrogram for the electrostatic potential of the $n=12$ simulation, $20000 < t \wci < 40000$.
  Some representations of the \Alfv{} continuum are overplotted. In thin white (curved) lines, the \Alfv{} continuum from the ideal kinetic model of LIGKA (neglecting pressure effects). In thick white dots, the continuum from the kinetic model of LIGKA, with the points being shown in large yellow dots if the imaginary part of the continuum $\Im(\omega)>0$. In thin (straight) white lines ($s>0.5$) and crosses, a simple model (implemented directly in the ORB5 diagnostics) for the continuum, with crossing points marking the analytical location of AEs.}.
  \label{fig:n12_spectrogram}
\end{figure}

If we perform the procedure described above on all radial points (independently), then we can stack the lower panels of figure~\ref{fig:n12_freq_growth} over radius, giving us a radial spectrogram, shown in figure~\ref{fig:n12_spectrogram}.
In this figure, we overlay the \Alfv{} continuum from LIGKA~\footnote{See discussion in~\ref{sec:ligka-q} regarding differences in the safety factor profile used for ORB5 and LIGKA.}~\cite{Lauber2009}, both with and without the pressure upshift (BAE gap).
%In this figure, we overlay the \Alfv{} continuum from LIGKA~\footnote{Although the equilibrium in LIGKA is not taken from CHEASE, but rather from HELENA~\cite{HELENA}. Since these equilibrium solvers have different boundary conditions at the magnetic axis, there is not an exact match of the q-profile. We have taken care to match the q-profile as closely as possible in the core. This difference is shown in figure~\ref{fig:ligka-q}.}~\cite{Lauber2009}, both with and without the pressure upshift (BAE gap).
This technique is designed to best identify frequencies, and we note that, since every radial position is normalized in a different way (to account for the growth/decay of the signal), a direct comparison of different radial positions should not be attempted.

An alternative method to what we have described above, which can also isolate growth rates and frequencies from multiple components, is the DMusic algorithm~\cite{Kleiber2021}, in which multiple complex frequencies are fit to the signal.
However, where we have a single dominant component at a given radius, we appreciate the fast numerical speed and high intuitivity of the method that we have described.

Considering the analysis described here, we can say that the least damped modes for the $n=12$ case are a BAE/RSAE at $s \approx 0.45$; and a TAE at $s \approx 0.65$. In addition, we also observe a weak EAE in the core, and another high frequency mode (high frequency RSAE or NAE (Triangularity-induced \Alfv{} eigenmode)), located around $s=0.15$ and $s=0.24$ respectively.

\begin{figure}
  \centering
  \includegraphics[width=0.4\textwidth]{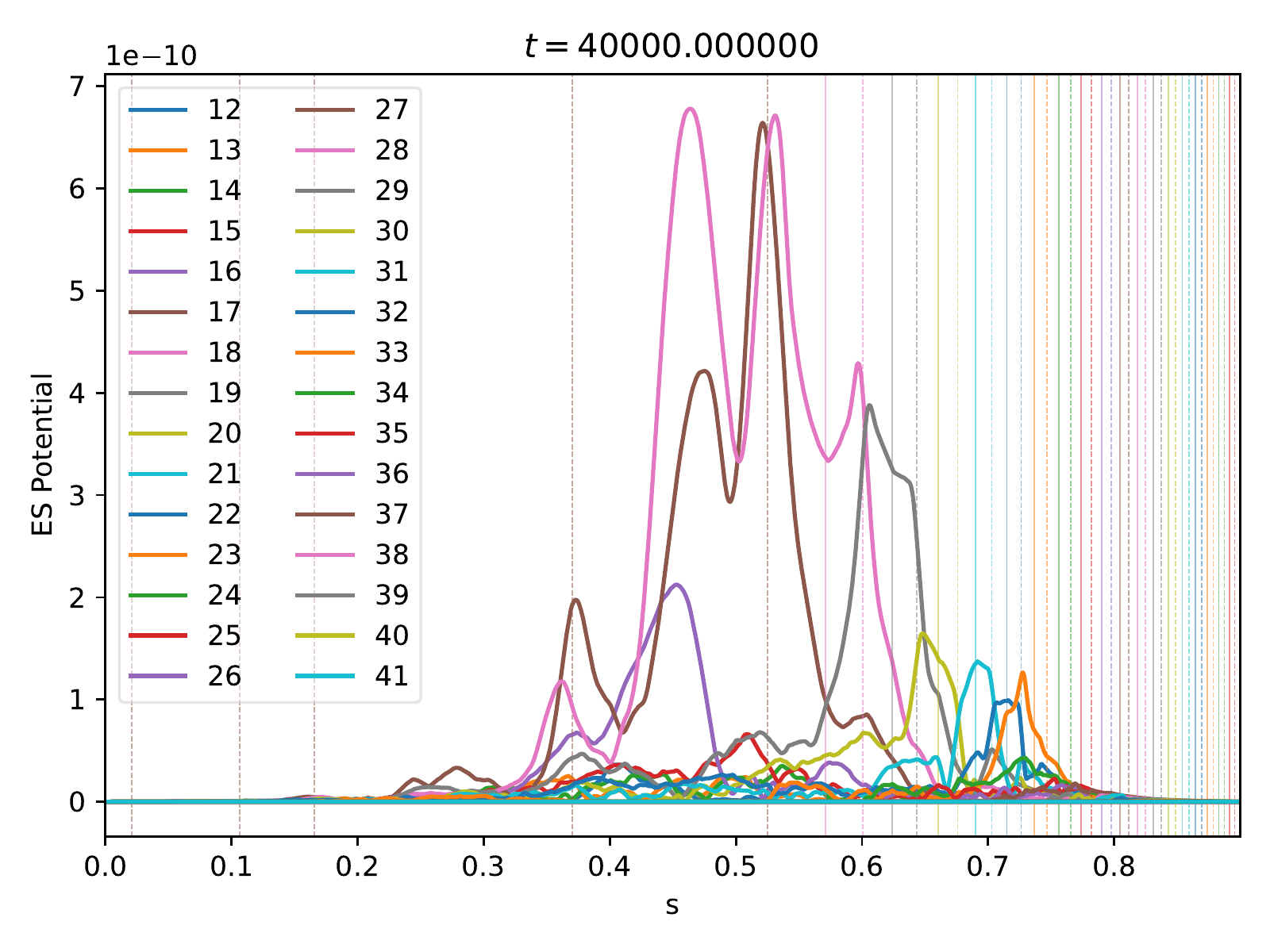}
  \caption{Plots of the absolute values of radial profiles of the $(n,m)=(16,m)$ helicities of the electrostatic potential, with the legend corresponding to $m$. Solid vertical lines of a given colour correspond to the location(s) of the $q=m/n$ rational surface associated with the $m$ in the legend; dashed vertical lines to the half-rational surfaces $q=(m+\frac{1}{2})/n$.}
  \label{fig:n16_m_of_s}
\end{figure}

\begin{figure}
  \centering
  \includegraphics[width=0.45\textwidth]{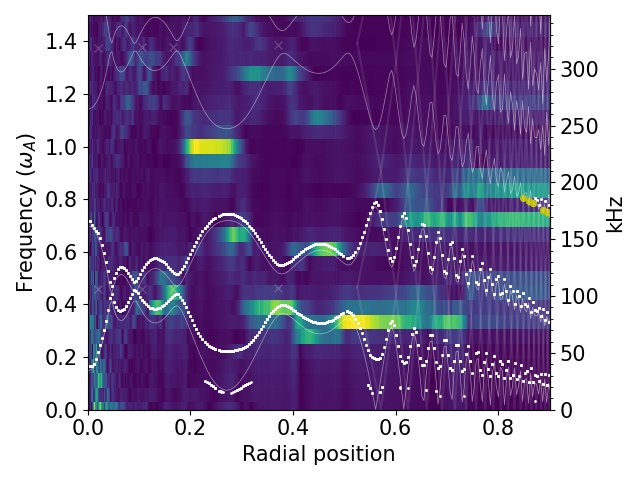}

  \caption{Radial frequency spectrogram for electrostatic potential of the $n=16$ simulation, $20000 < t \wci < 40000$.
%  b) Radial frequency spectrogram calculated using the DMusic algorithm.
  For the explanation of the overplotted \Alfv{} continuum, see figure~\ref{fig:n12_spectrogram}.}
  \label{fig:n16_spectrogram}
\end{figure}

Next, we consider a simulation with $n=16$.
Similar to the analysis for $n=12$, we show in figure~\ref{fig:n16_m_of_s} the radial profiles of the poloidal harmonics of the electrostatic potential $\Phi$.
Here, we also mark with vertical lines the rational surfaces (solid) and half-rational surfaces (dashed).
In figure~\ref{fig:n16_spectrogram}, we again show the radial spectrogram of the analysis.
Here we see some TAEs ($s=(0.37, 0.52)$), with an RSAE at the minimum of q around $s=0.46$. We also observe a very weak high frequency mode in the EAE gap  at $s \approx 0.25$.
Again, the observed damping is weak, with the measured fit damping rates $-0.005 < \gamma / \wa \leq 0$ across a broad range of the plasma, in particular the region $0.4 \leq s \leq 0.6$.

\begin{figure}
  \centering
  \includegraphics[width=0.4\textwidth]{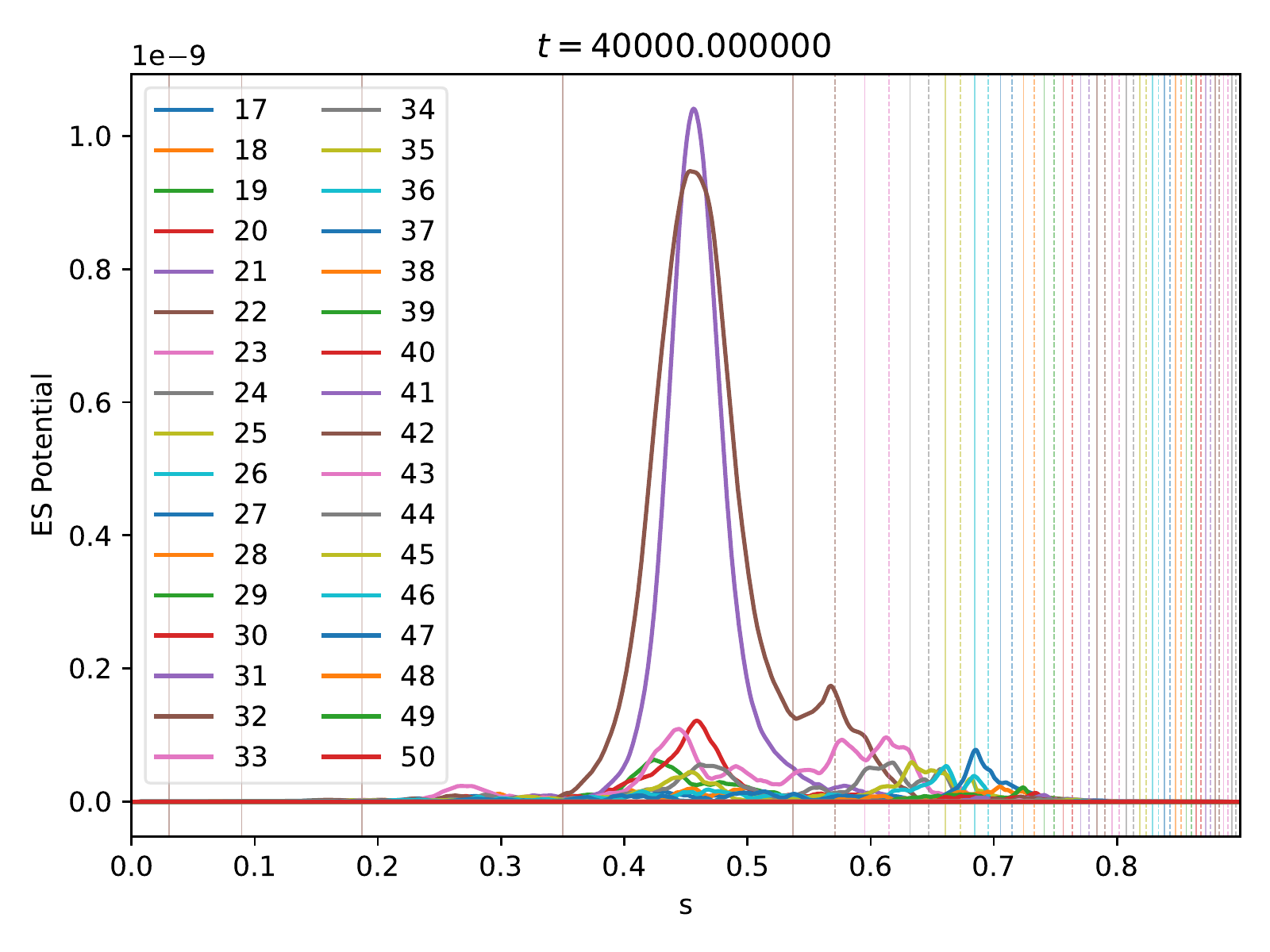}
  \caption{Plots of the absolute values of radial profiles of the $(n,m)=(20,m)$ helicities of the electrostatic potential, with the legend corresponding to $m$.}
  \label{fig:n20_m_of_s}
\end{figure}

\begin{figure}
  \centering
  \includegraphics[width=0.45\textwidth]{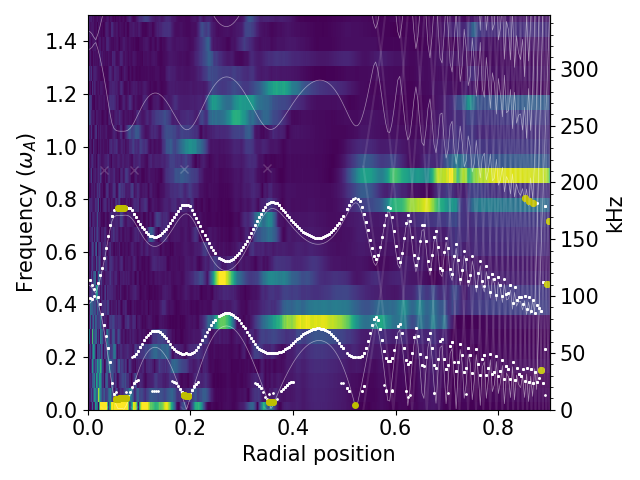}
  \caption{Radial frequency spectrogram for electrostatic potential of the $n=20$ simulation, $20000 < t \wci < 40000$.
  For the explanation of the overplotted \Alfv{} continuum, see figure~\ref{fig:n12_spectrogram}.}
  \label{fig:n20_spectrogram}
\end{figure}

\begin{figure}
  \centering
  \includegraphics[width=0.35\textwidth]{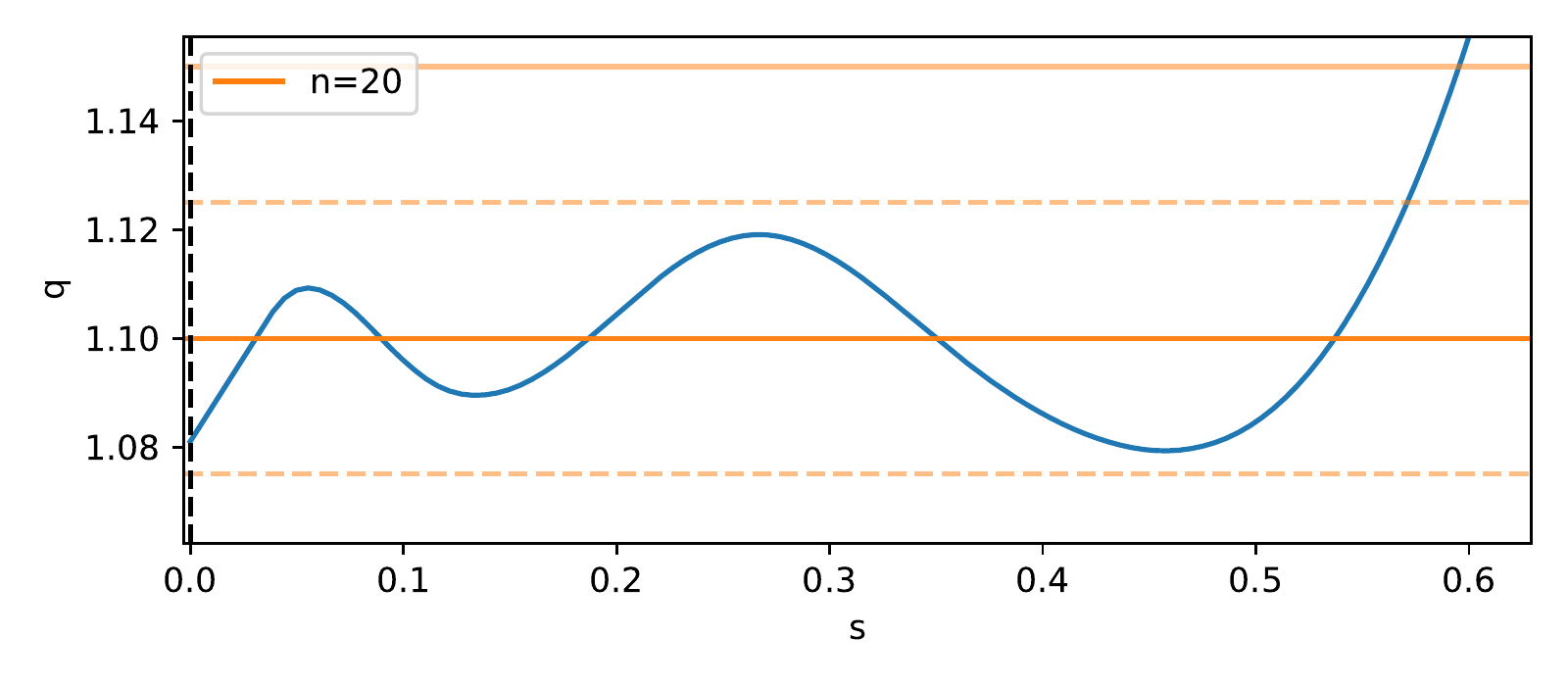}
  \caption{Safety factor profile in core of the plasma with the $n=20$ rational (solid) and the half-rational (dashed) surfaces marked.}
  \label{fig:n20_q}
\end{figure}

Next we consider $n=20$, for which we increase $(N_s, N_{\theta^*}, N_\varphi) = (1024, 512, 128)$.
The radial profile of the electrostatic potential is show in figure~\ref{fig:n20_m_of_s}, showing a peak of two poloidal harmonics ($m=20$ and $21$), peaked at approximately $s=0.45$.
We note that, even though this modes structure strongly reminds us of a TAE, we do not have a half-rational surface at $s=0.45$, located between two rational surfaces ($s=0.350$, $s=0.537$), but not on a half-rational surface.
We show in figure~\ref{fig:n20_q} that the minimum of the safety factor in this region is very close to the half-rational value of a TAE.
From the spectrogram in figure~\ref{fig:n20_spectrogram}, we note that the frequency of this mode is close to the bottom of the TAE gap, similar to behaviour also observed in the ramp-up phase of tokamaks~\cite{Lauber2013, Van_Zeeland2016}
We therefore label this mode an TAE-like RSAE.
We observe the weakest growth in the region $0.3 \leq s \leq 0.5$, where $|\gamma|<$~\SI{0.005}{\wa}, where the damping is weak and the mode is close to marginal stability.

\begin{figure}
  \centering
  \includegraphics[width=0.4\textwidth]{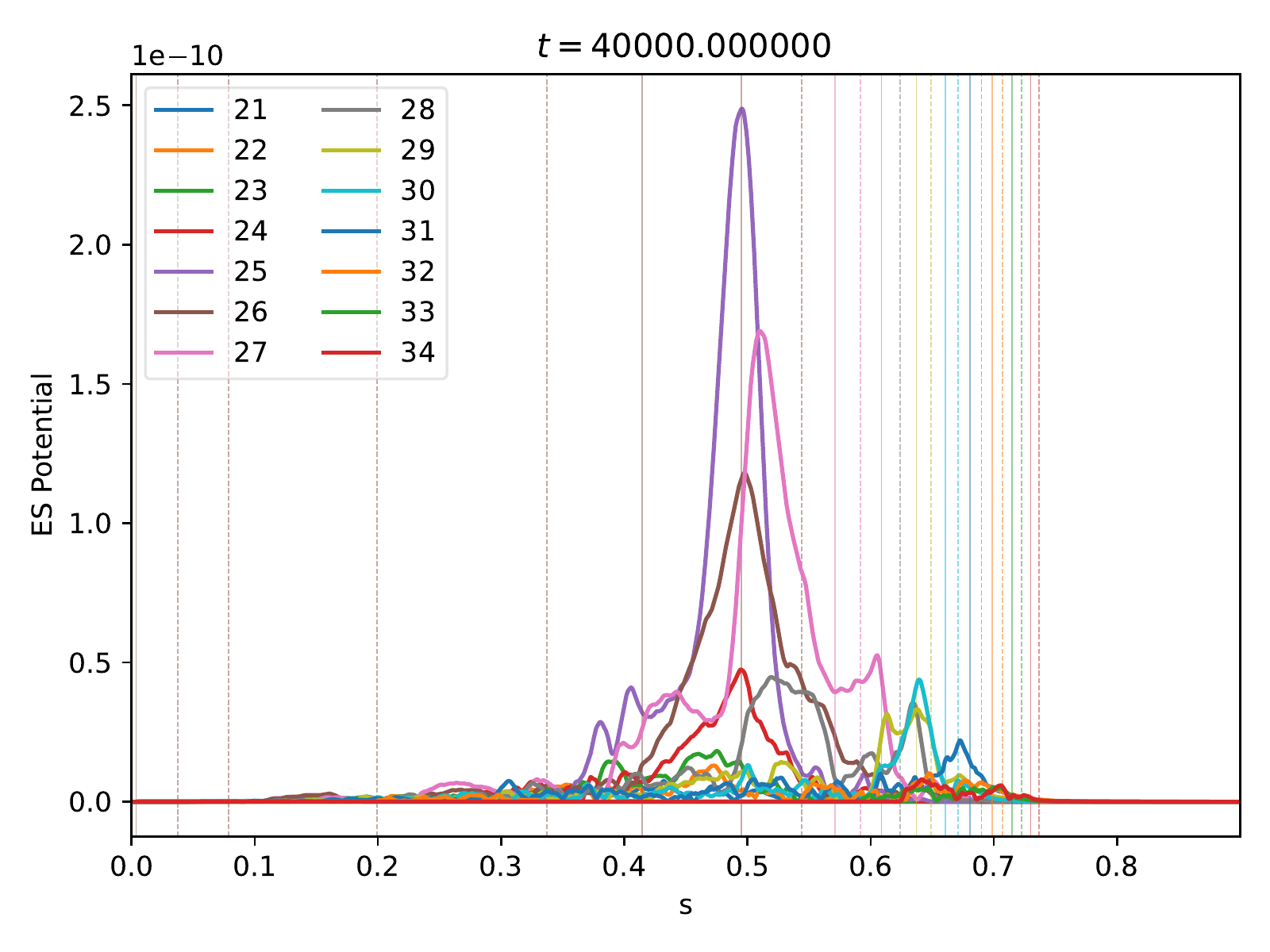}
  \caption{Plots of the absolute values of radial profiles of the $(n,m)=(24,m)$ helicities of the electrostatic potential, with the legend corresponding to $m$. Since the number of poloidal harmonics in the simulation is large ($21 \leq m \leq 59$), we show only $m \leq n+10$.}
  \label{fig:n24_m_of_s}
\end{figure}

\begin{figure}
  \centering
  \includegraphics[width=0.45\textwidth]{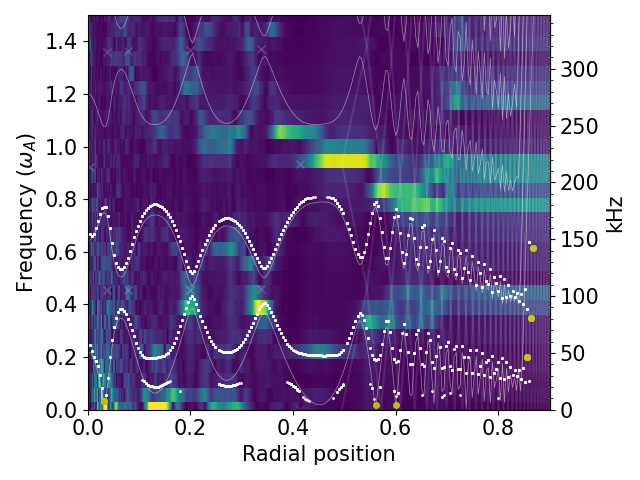}
  \caption{Radial frequency spectrogram for electrostatic potential of the $n=24$ simulation, $20000 < t \wci < 40000$.
  For the explanation of the overplotted \Alfv{} continuum, see figure~\ref{fig:n12_spectrogram}.}
  \label{fig:n24_spectrogram}
\end{figure}

For $n=24$, from figures~\ref{fig:n24_m_of_s}~and~\ref{fig:n24_spectrogram}, we find the predominant modes are found in the EAE band, distributed between $s=0.4$ and $0.6$.

\begin{figure}
  \centering
  \includegraphics[width=0.4\textwidth]{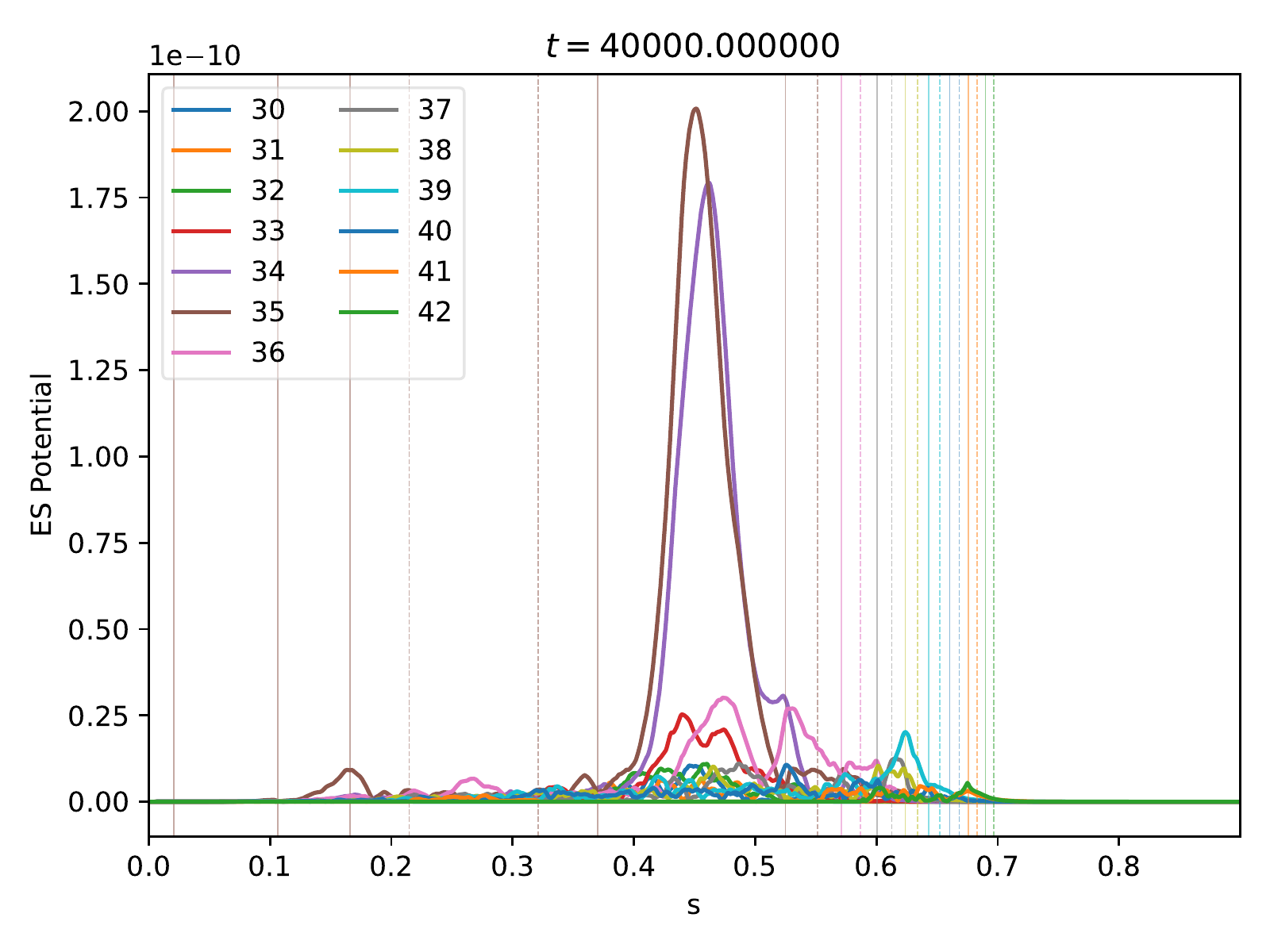}
  \caption{Plots of the absolute values of radial profiles of the $(n,m)=(32,m)$ helicities of the electrostatic potential, with the legend corresponding to $m$. Since the number of poloidal harmonics in the simulation is large ($30 \leq m \leq 77$), we show only $m \leq n+10$.}
  \label{fig:n32_m_of_s}
\end{figure}

For $n=32$, like the case of $n=20$, we see a narrow TAE-like mode structure in figure~\ref{fig:n32_m_of_s}, with the dominant poloidal harmonics $m=(34,35)$.
Once again in the frequency, we see that the mode is in the gap of the minimum of the safety profile, with a frequency close to the TAE frequency.
Since $(21+\frac{1}{2})/20$ and $(34+\frac{1}{2})/32$ are very similar, 1.075 and 1.078125 respectively, we can compare to figure~\ref{fig:n20_q}, and apply the same arguments regarding the proximity of $q_\textrm{TAE}$ and the $q_\textrm{min}$.

\begin{figure}
  \centering
  \includegraphics[width=0.45\textwidth]{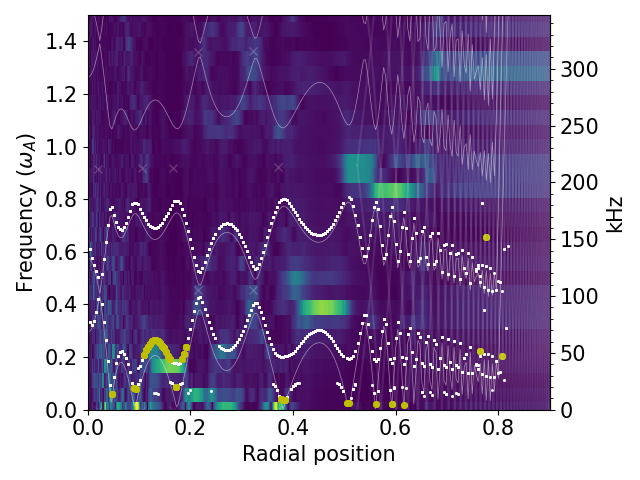}
  \caption{Radial frequency spectrogram for electrostatic potential of the $n=32$ simulation, $20000 < t \wci < 40000$.
  For the explanation of the overplotted \Alfv{} continuum, see figure~\ref{fig:n12_spectrogram}.}
  \label{fig:n32_spectrogram}
\end{figure}

\subsection{Unstable mesoscale modes}
\label{sec:results_aitg}
Further increasing the toroidal mode number, we observe a transition to a very localized instability close to the $q_\textrm{max}$ at $s = 0.27$.
We therefore present analysis for simulations performed on a reduced radial domain, $0.1 \leq s \leq 0.4$, which retain the physical effects, but allows us to reduce the numerical cost for such simulations (both in terms of reducing the required radial resolution as well as the $q_\textrm{max}$, and therefore the poloidal resolution required to resolve $m \approx n q_\textrm{max}$.
We therefore run a simulation for $n=50$ with, $(N_s, N_{\theta^*}, N_\varphi) = (384, 480, 240$).
This simulation has been performed with the number of markers, $N_p = (64,64,16,16)\times 10^6$, for H, $e^{-}$, Be, Ne.

\begin{figure}
  \centering
  \includegraphics[width=0.4\textwidth]{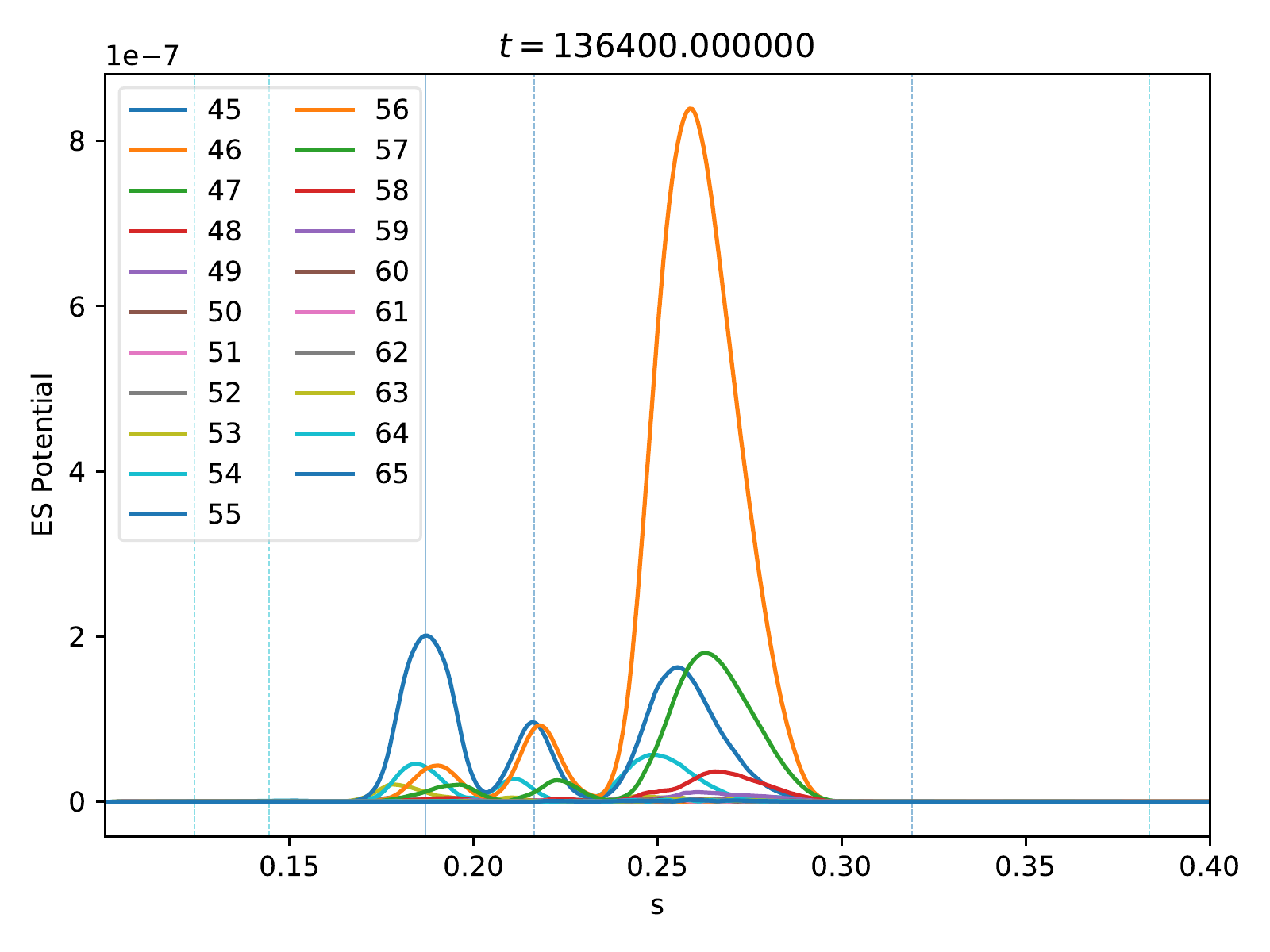}
  \caption{Plots of the absolute values of radial profiles of the $(n,m)=(50,m)$ helicities of the electrostatic potential, with the legend corresponding to $m$. Vertical lines correspond to the locations of rational and half-rational surfaces, see figure~\ref{fig:n16_m_of_s}.}
  \label{fig:n50_m_of_s}
\end{figure}

\begin{figure}
  \centering
  \includegraphics[width=0.45\textwidth]{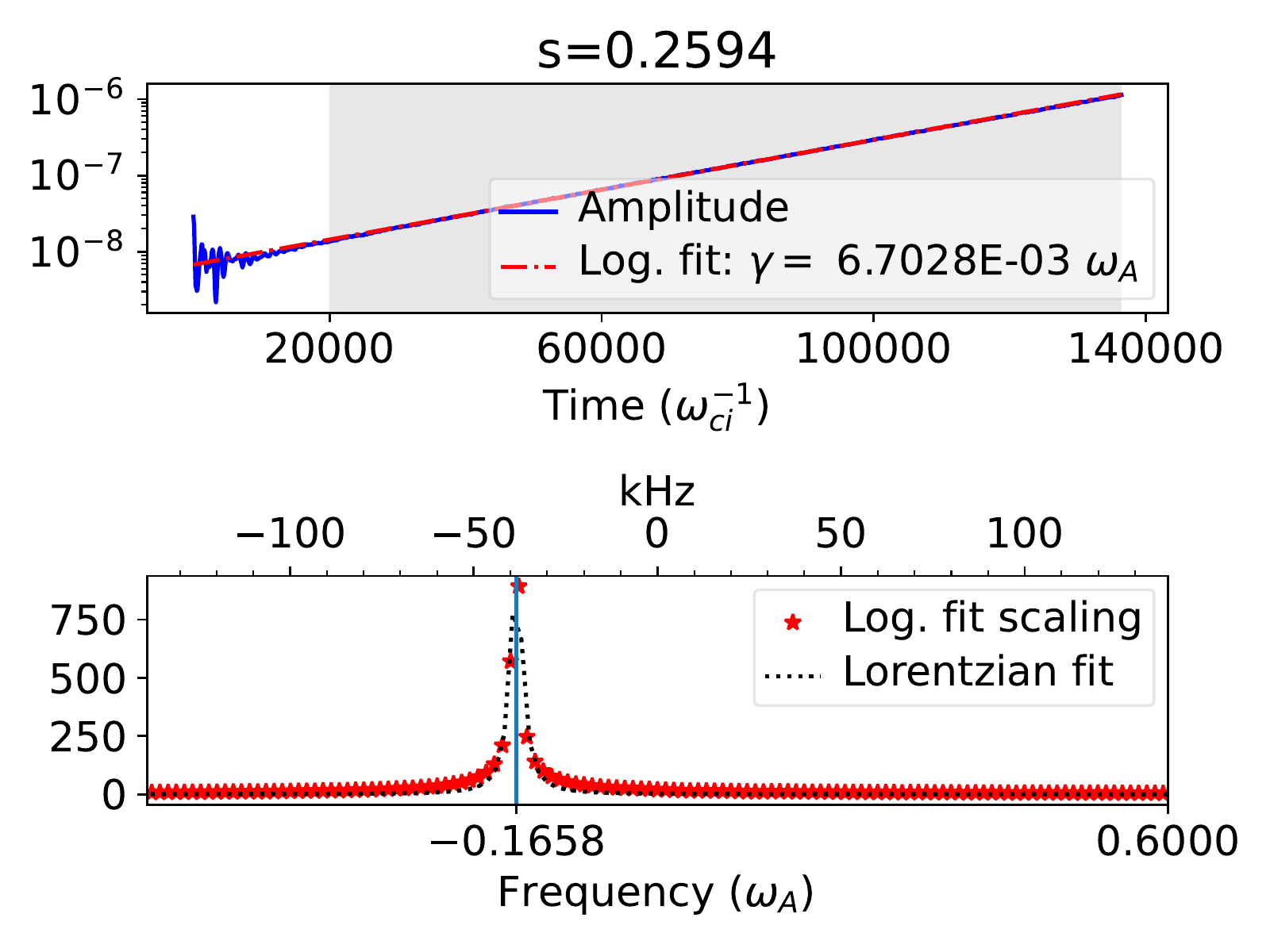}
  \caption{ For the point $s=0.259$: (a) Plot of the $n=50$ envelope, with an exponential fit to the extrema in the range $20000 < t \wci$ (shaded grey). (b) Discrete Fourier transform of the electrostatic potential after dividing by the fitted exponential in (a).}
  \label{fig:n50_freq_growth}
\end{figure}

In figure~\ref{fig:n50_m_of_s}, we can see that the dominant mode is single poloidal harmonic ($m=56$) peaked at $s=0.26$.
We also see a secondary peak of $m=55$ peaked at the rational surface at $s=0.187$.
In figure~\ref{fig:n50_freq_growth}, we see that the mode is growing with a clearly identified growth rate and frequency, ($\omega$, $\gamma$) = ($0.1658$, $6.7 \times 10^{-3}$) $\wa$, or $f=$\SI{38.4}{\kilo \hertz} ($\gamma/\omega=$~\SI{4.0}{\%}).

\begin{figure}
  \centering
  \includegraphics[width=0.4\textwidth]{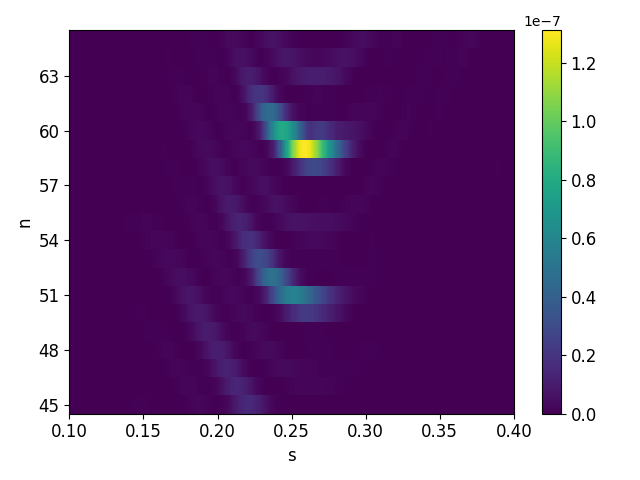}
  \caption{Snapshots of the electrostatic potential on the outboard midplane for 21 simulations with $n=\{45,\ldots,65\}$ at $t\wci=40000$, plotted against the radius and the toroidal mode number.}
  \label{fig:lown_Afinal_of_ns}
\end{figure}

If we extend this case to include other toroidal mode numbers, we find that after a fixed amount of time, the electrostatic potential on outboard midplane can be visualized in figure~\ref{fig:lown_Afinal_of_ns}.
Here we see a pattern with a near periodicity of $\Delta n$ of approximately 8.
We can understand this pattern by noting that the local $q_\textrm{max} = 1.1191$, and therefore the proximity of the rational surfaces to the $q_\textrm{max}$ is periodic with $1/(1-q_\textrm{max}) = 8.4$.
This also explains why a peak in intensity is observed for $n=59$, since the rational fraction $66/59$ is the closest approximation of $\qmax$ in this range of $n$ in the denominator.

\begin{figure}
  \centering
  \includegraphics[width=0.3\textwidth]{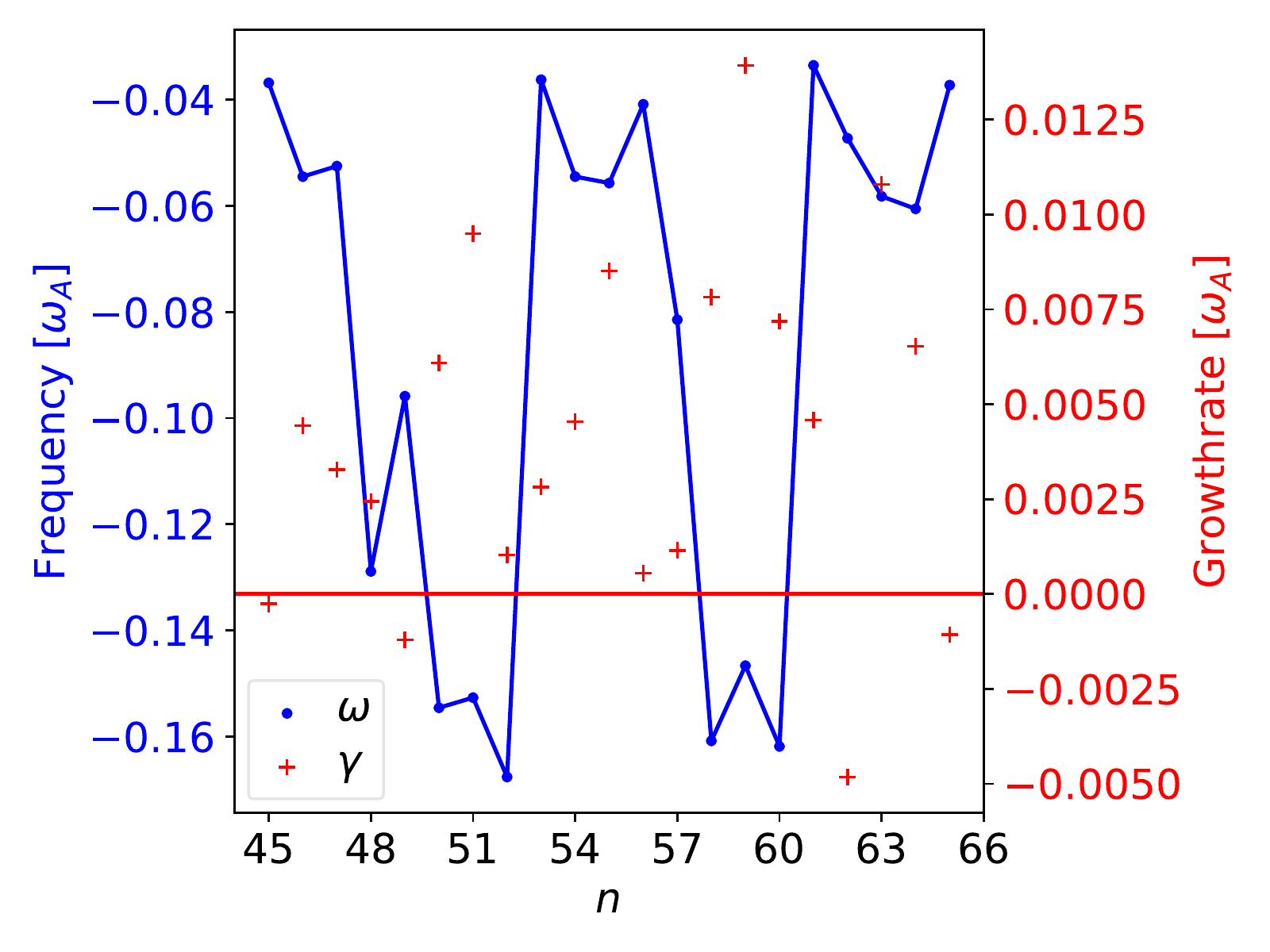} \includegraphics[width=0.3\textwidth]{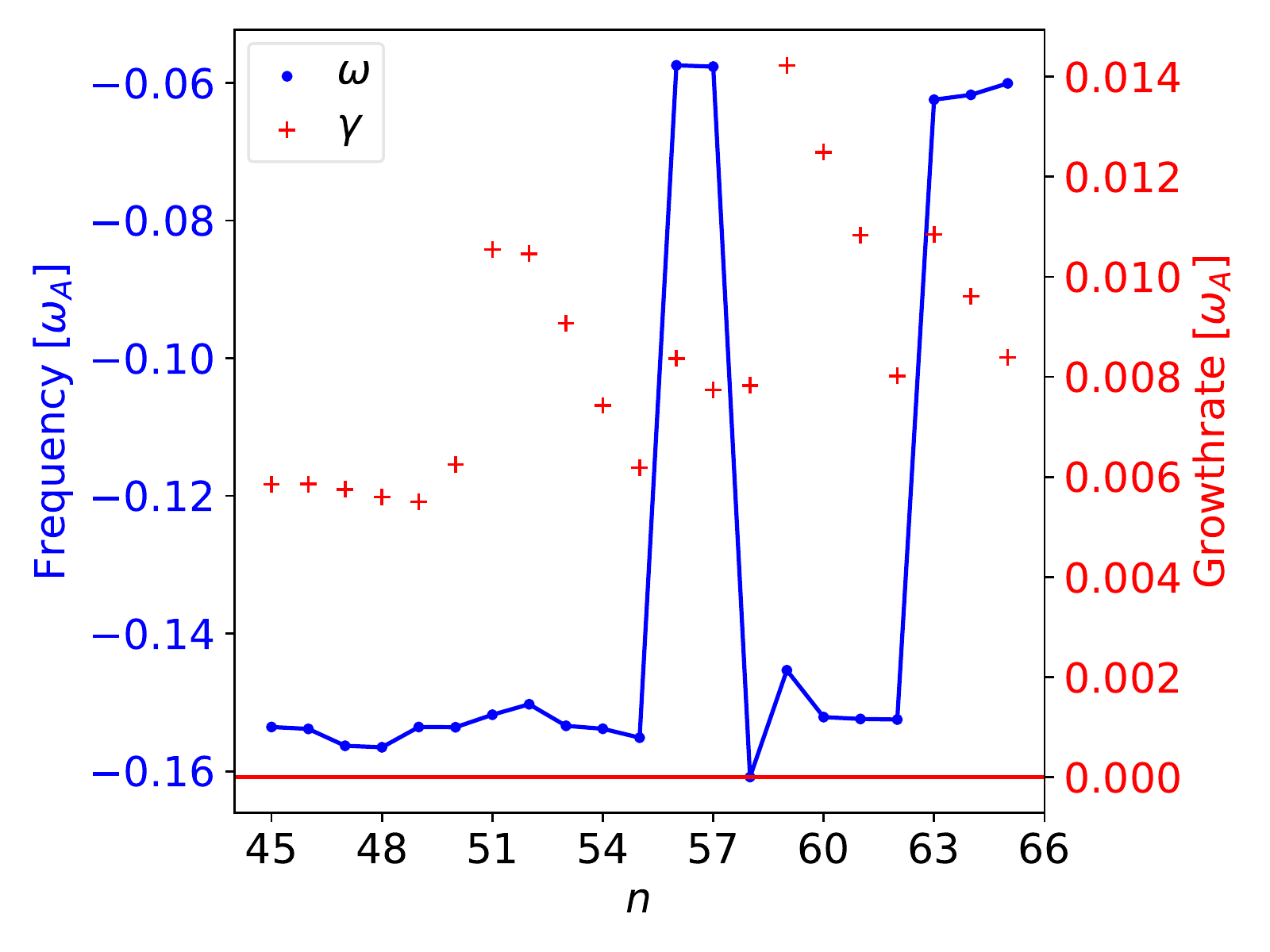}
  \caption{Frequency (blue lines) and growth rate (red points) for different mode numbers fit at the radial position of (a) the maximum of $q$ at $s=0.265$, (b) the location of the respective peak instability.}
  \label{fig:lown_gamma_of_n}
\end{figure}

In figure~\ref{fig:lown_gamma_of_n}, we show the frequency and growth rate for these cases, both located at the $\qmax$ (a), and at the location of the mode peak (b).
Here we can further see the almost periodic behaviour in the growth and frequency.

\begin{figure}
  \centering
  \includegraphics[width=0.45\textwidth]{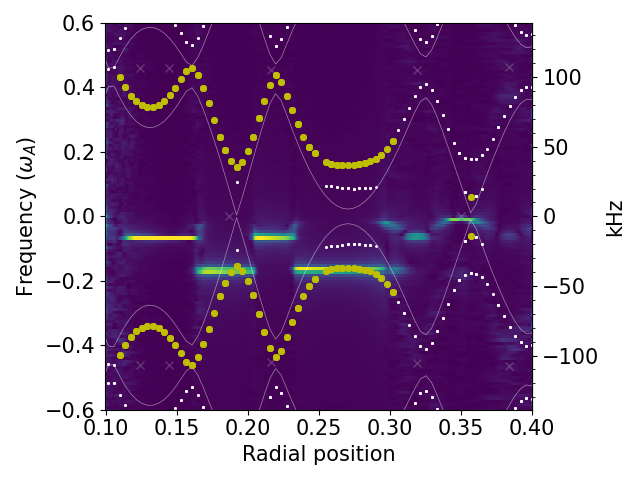}
  \caption{Radial frequency spectrogram for electrostatic potential of the $n=50$ simulation, $20000 < t \wci$.
  For the explanation of the overplotted \Alfv{} continuum, see figure~\ref{fig:n12_spectrogram}.
  The Fourier analysis has been performed on the complex spatial Fourier coefficients, and therefore the sign of the frequency can be determined.
  In this case, the LIGKA data has been mirrored in the $\Im(\omega)=0$ plane.}
  \label{fig:n50_spectrogram}
\end{figure}

Regarding the mode drive, we show in figure~\ref{fig:n50_spectrogram} the spectrogram, but we draw particular attention to the kinetic spectrum overplotted from LIGKA.
This is obtained from the local kinetic model of LIGKA, and we have plotted in yellow the points where the imaginary part of the continuum frequency is positive.
Since the mode that we observe corresponds with tip of this unstable continuum, and since the mode is close to the position where the $\qmax$ is close to the value of a rational fraction, we label this mode a BAE~\cite{Heidbrink2008}, or \Alfv{}ic Ion Temperature Gradient (AITG) mode~\cite{Zonca1996, Zonca1998}, although we note that the mode also has similarities with the RSAE.
We find ($\omega$, $\gamma$) = $(-0.152, 6.15 \times 10^{-3})$ $\omega_A$.

%[0.2593750059604645]
%frequency_wci = -0.0009323695263924099
%frequency_wa = -0.15217970847667087
%frequency_rads = -241166.70169666075
%frequency_hz = -38382.87268419214
%lorentz_method = 3
%growthrate3_wa = 0.0061534836813511695
%growthrate3_frequency = 0.04043563851546277
%tfrom = 20000.0
%tto = 136400.0
%spos = 0.2593750059604645

Wanting to be sure that the artificial electron mass is not responsible for the mode growth in the simulations, we perform also a simulation with $m_\trm{H}/m_\trm{e} = 1600$ (close to nominal).
We therefore reduce the time step to $\Delta t =$ \SI{1.5}{\wci^{-1}}, and obtain values of ($\omega$, $\gamma$) = $(-0.169, 7.93 \times 10^{-3})$ $\omega_A$.

% mr1600, dt1.5
%[0.2601562440395355]
%frequency_wci = -0.0009509724836960729
%frequency_wa = -0.1690719199170261
%frequency_rads = -245978.54263282623
%frequency_hz = -39148.70095455481
%lorentz_method = 3
%growthrate3_wa = 0.007927753262297792
%growthrate3_frequency = 0.04688982810503615
%tfrom = 20000.0
%tto = 58875.0
%spos = 0.2601562440395355

Further decreasing the time step to $\Delta t =$~\SI{0.5}{\wci^{-1}}, we find ($\omega$, $\gamma$) = $(-0.161, 8.91 \times 10^{-3})$~$\omega_A$, or $f =$~\SI{-37.4}{\kilo \hertz}~($\gamma/|\omega| =$~\SI{5.5}{\%}).
We therefore have confidence that this mode is driven unstable by the bulk plasma in the model considered, and not that this is a numerical instability.
For a detailed discussion of the effect of the electron mass and the time step on the numerical convergence, we refer to Reference~\cite{Hayward-Schneider2021}.

% mr1600, dt0.5
%[0.2601562440395355]
%frequency_wci = -0.0009092427061138526
%frequency_wa = -0.16165284761525453
%frequency_rads = -235184.71836340913
%frequency_hz = -37430.81046721181
%lorentz_method = 3
%growthrate3_wa = 0.008911930533353252
%growthrate3_frequency = 0.05513005594905629
%tfrom = 20000.0
%tto = 48750.0
%spos = 0.2601562440395355

\subsubsection{Effect of energetic particles on BAE/AITG modes}
\label{sec:results_aitg_ep}
\begin{figure}
  \centering
  \includegraphics[width=0.39\textwidth]{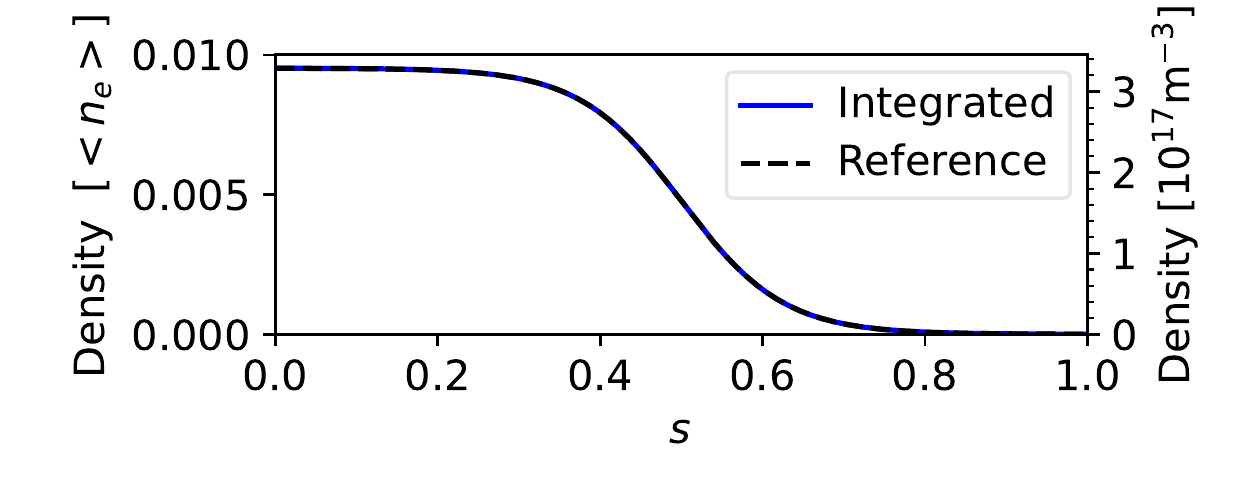}
  \caption{Energetic particle density profile, obtained by integrating the moments of the distribution function output from a simulation.}
  \label{fig:iter_n_nbi}
\end{figure}

An interesting question is whether energetic particles might have any effect on such a mode such as this bulk-plasma-driven mode.
Still with the $m_\trm{H}/m_\trm{e}$ and $\Delta t =$\SI{0.5}{\wci^{-1}}, we therefore add a simplified population of NBI particles, using an isotropic slowing down distribution with an injection energy of \SI{1}{\mega \electronvolt}.
The effective temperature of this distribution depends weakly on radius and is shown in figure~\ref{fig:iter_T_nbi}.
We consider a simplified density profile for the NBI, shown in figure~\ref{fig:iter_n_nbi}, constructed with a $tanh$ function.

With the addition of the NBI, we find ($\omega$, $\gamma$) = $(-0.164, 8.88 \times 10^{-3})$~$\omega_A$, or $f =$~\SI{-38.1}{\kilo \hertz} ($\gamma/|\omega|=$~\SI{5.4}{\%}).

We therefore state that energetic particles, at least those considered here, have no effect on the linear behaviour of these BAE/AITG modes.

%[0.26093751192092896]
%frequency_wci = -0.0009243799205915033
%frequency_wa = -0.164344069451648
%frequency_rads = -239100.11026019827
%frequency_hz = -38053.96444172776
%lorentz_method = 3
%growthrate3_wa = 0.008883480870625076
%growthrate3_frequency = 0.05405416149341916
%tfrom = 20000.0
%tto = 39975.0
%spos = 0.26093751192092896

\subsection{Microscale instabilities}
\label{sec:results_micro}
Considering now the microscales, we perform a study of large toroidal mode numbers, looking for ion gyroradius-scale micro-instabilities (such as those associated with turbulence).

We consider the radial domain $0.4 \leq n \leq 0.7$, the region with the steepest electron temperature gradient and consider toroidal mode numbers in the range $n>120$, increasing the toroidal resolution to keep $N_{\varphi} > 3 n$, ideally $\geq4 n$, but also choosing convenient values for parallelization.

We begin below by showing the case of $n=180$, with $(N_s, N_{\theta^*}, N_\varphi) = (384, 1152, 576)$.
In the late linear phase, we find a peak at $s=0.467$, from where we measure low or zero frequency, and a growth rate of $\gamma=$~\SI{4.15e5}{\per\second}.

In the case of a larger mode number, $n=216$, we find that the mode is linearly stable.
For a smaller mode number, $n=160$, we find a much weaker growth rate, which allows us to measure more accurately the linear (non-zero) frequency, finding ($5000 \leq t\wci^{-1} \leq 54300$) $\omega =$~\SI{-4.7}{\kilo \hertz}, $\gamma =$~\SI{5.9432e4}{\per\second}.

We have therefore identified this range of toroidal mode numbers, $n < 216$, with a peak of $\gamma$ for $n\approx 180$, which driven unstable by bulk-plasma gradients.

% Changing the $T_e$ profile to match that of $T_i$, we find a slight change in the growth rate to $\gamma=$~\SI{4.34e5}{\per\second}.

\begin{figure}
  \centering
  \includegraphics[width=0.45\textwidth]{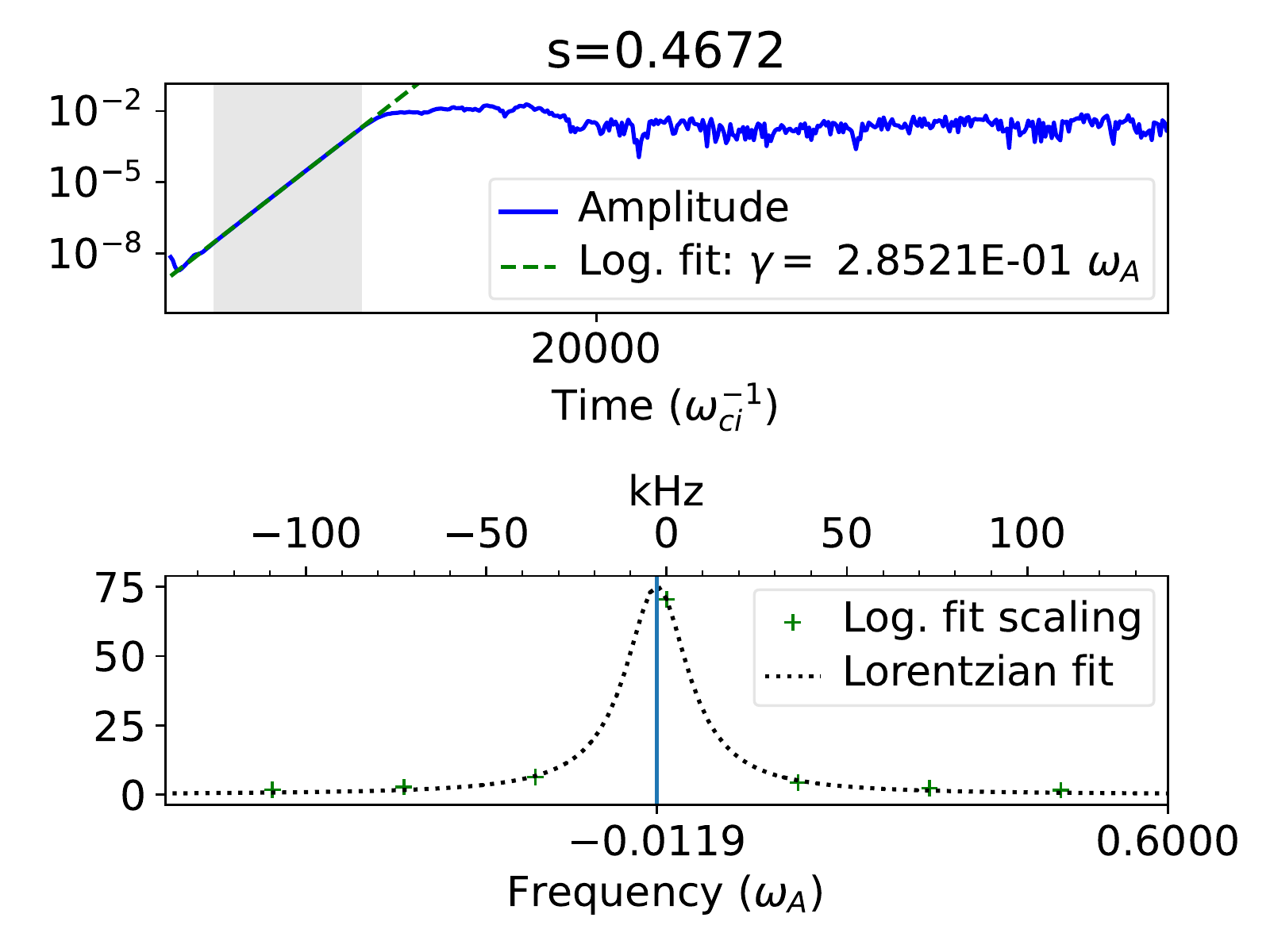}
  \caption{ For the point $s=0.4672$: (a) Plot of the $n=180$ envelope, with an exponential fit to the extrema in the range $2000 < t \wci < 9000$ (shaded grey). (b) Discrete Fourier transform of the electrostatic potential after dividing by the fitted exponential in (a).}
  \label{fig:n180_freq_growth}
\end{figure}

% n180 nominal
%[0.4671874940395355]
%frequency_wci = -6.666509402609118e-05
%frequency_wa = -0.011852283458963397
%frequency_rads = -17243.593220788745
%frequency_hz = -2744.4030977544253
%lorentz_method = 2
%growthrate2_per_s = 414945.1237555052
%growthrate2_wa = 0.28521011622657244
%growthrate2_frequency = 24.06372723147114
%tfrom = 2000.0
%tto = 9000.0
%spos = 0.4671874940395355

% n180 TeTi
%[0.46953123807907104]
%frequency_wci = -9.027279391963466e-05
%frequency_wa = -0.016049459732995323
%frequency_rads = -23349.960875252702
%frequency_hz = -3716.2616942988266
%lorentz_method = 2
%growthrate2_per_s = 433897.5793804924
%growthrate2_wa = 0.2982369763151044
%growthrate2_frequency = 18.582368582910807
%tfrom = 2000.0
%tto = 9000.0
%spos = 0.46953123807907104

% n160
%[0.48828125]
%frequency_wci = -0.00011555891226461787
%frequency_wa = -0.020545039414985127
%frequency_rads = -29890.46824636606
%frequency_hz = -4757.215772740496
%lorentz_method = 2
%growthrate2_per_s = 59432.97914499556
%growthrate2_wa = 0.04085091237178503
%growthrate2_frequency = 1.9883589194766509
%tfrom = 5000.0
%tto = 54300.0
%spos = 0.48828125

%% file: conclusions.tex
\section{Summary and outlook}
\label{sec:conc}
In conclusion, we have shown a multi-scale nature of instabilities in this ITER PFPO scenario.
By means of global electromagnetic gyrokinetic simulations with ORB5, we have identified three classes of eigenmode/instability, namely the weakly damped `macroscale' \Alfv{} eigenmodes in the range $n = [10\to35]$, a range of `mesoscale' unstable bulk-plasma driven \Alfv{}ic instabilities in the range $n=[45\to60]$, and a range of unstable `microscale' instabilities in the range $n=[150\to220]$.

We have shown that, for an isotropic slowing down distribution function of NBI energetic particles with $E_\trm{birth} =$~\SI{1}{\mega \electronvolt}, the growth rates of these mesoscale modes appear insensitive to the addition of energetic particles, although a more realistic treatment of the EP distribution function is left for future work.

We have identified a number of effects (that we refer to as `multi-scale') which would be of strong interest to consider in a unified setting.
Since there are partially different radial locations key to the different single-scale phenomena, the nonlinear interplay of these different scales would interesting.
Considering the numerical demands of dealing with large mode numbers, and time scale separation and radial domains, this would be very demanding, however we think that we have laid out the scales and phenomena which such an effort should retain.

Finally, we note that the label `macroscale' is often used to refer to MHD instabilities, for example fishbone and sawtooth physics, which are not included in this study, but are also relevant for this scenario.
We note that the safety factor profile, $q(s)$, is of major importance for the \Alfv{}ic modes, and any sawtooth activity will lead to a change in this profile.
It is therefore of particular interest to consider time-dependent scenarios, which should especially be considered with reduced models~\cite{Popa_MSc}.
The study of these MHD phenomena (in isolation) is of interest from (gyro-)kinetic codes or (kinetic-) MHD models, especially since these might lead to a redistibution of the energetic particle profile.

%% file: acknowledgements.tex
\section*{Acknowledgements}
The authors would like to acknowledge A.~Polevoi for providing the scenario data through IMAS, and to S.~Pinches and M.~Schneider for discussions relating to the scenario and to dealing with IMAS data.
The authors would like to acknowledge discussions with those participating in the ITPA-EP activity B.11.12, and the associated collaboration with the ISEP SciDAC project.
% TODO:...
The authors would also like to thank the members of the ORB5 team.
% ITPA, ISEP, ITER
% This work has been carried out within the framework of the EUROfusion Consortium and has received funding from the Euratom research and training program 2014-2018 and 2019-2020 under grant agreement No 633053. The views and opinions expressed herein do not necessarily reflect those of the European Commision. %FP8-only
This work has been carried out within the framework of the EUROfusion Consortium, funded by the European Union via the Euratom Research and Training Programme (Grant Agreement No 101052200 -- EUROfusion). Views and opinions expressed are however those of the author(s) only and do not necessarily reflect those of the European Union or the European Commission. Neither the European Union nor the European Commission can be held responsible for them. %FP9 or 8&9
Simulations presented in this work were performed on the MARCONI FUSION HPC system at CINECA and the HPC systems of the Max Planck Computing and Data Facility (MPCDF).

%% file: appendix.tex
\appendix
\section{Comparison of safety factor profile}
\label{sec:ligka-q}
\begin{figure}
  \centering
  \includegraphics[width=0.45\textwidth]{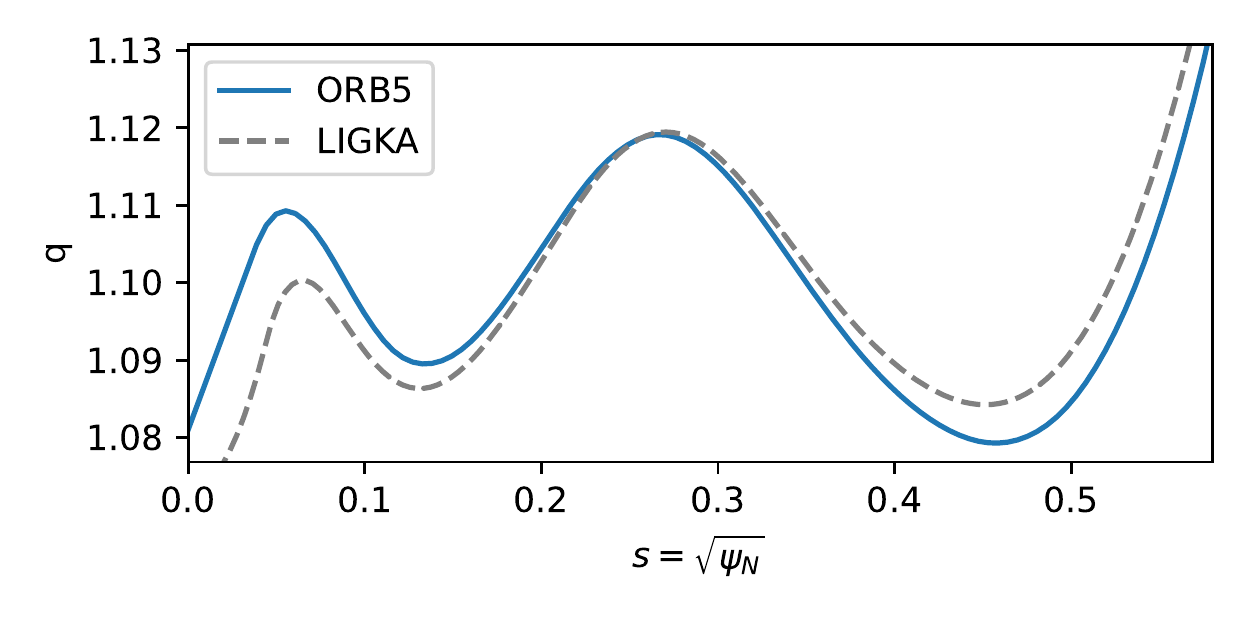}
  \caption{Comparison of the safety factor profiles (ORB5, from CHEASE; LIGKA, from HELENA) in the core of the plasma. The ``ORB5'' line corresponds to the inset plot in figure~\ref{fig:iter_eq}(b).}
  \label{fig:ligka-q}
\end{figure}

As noted previously, when we overlay the \Alfv{} continuum from LIGKA, we must note that the safety factor profile is not consistent between LIGKA and ORB5, since these two codes interface to different equilibrium solvers, HELENA~\cite{HELENA} and CHEASE respectively.
Since these equilibrium solvers have different boundary conditions at the magnetic axis, there is not an exact match of the q-profile.
We have taken care to match the q-profile as closely as possible in the core region, especially to match the region relevant for \S\ref{sec:results_aitg}.
We therefore plot a comparison of the safety factor profiles in figure~\ref{fig:ligka-q}.

%% file: twhs_nf22.bbl
\begin{thebibliography}{10}

\bibitem{Lanti2020}
E.~{Lanti}, N.~{Ohana}, N.~{Tronko}, T.~{Hayward-Schneider}, A.~{Bottino},
  B.~F. {McMillan}, A.~{Mishchenko}, A.~{Scheinberg}, A.~{Biancalani},
  P.~{Angelino}, S.~{Brunner}, J.~{Dominski}, P.~{Donnel}, C.~{Gheller},
  R.~{Hatzky}, A.~{Jocksch}, S.~{Jolliet}, Z.~X. {Lu}, J.~P. {Martin Collar},
  I.~{Novikau}, E.~{Sonnendr{\"u}cker}, T.~{Vernay}, and L.~{Villard}.
\newblock {ORB5: a global electromagnetic gyrokinetic code using the PIC
  approach in toroidal geometry}.
\newblock {\em Computer Physics Communications}, 251, 2020.

\bibitem{Mishchenko2019}
A.~Mishchenko, A.~Bottino, A.~Biancalani, R.~Hatzky, T.~Hayward-Schneider,
  N.~Ohana, E.~Lanti, S.~Brunner, L.~Villard, M.~Borchardt, R.~Kleiber, and
  A.~K{\"o}nies.
\newblock Pullback scheme implementation in {ORB5}.
\newblock {\em Computer Physics Communications}, 238, 2019.

\bibitem{Ligka}
Ph. Lauber, S.~G{\"u}nter, A.~K{\"o}nies, and S.D. Pinches.
\newblock {LIGKA}: A linear gyrokinetic code for the description of background
  kinetic and fast particle effects on the {MHD} stability in tokamaks.
\newblock {\em Journal of Computational Physics}, 226(1), 2007.

\bibitem{McMillan2010}
B.F. McMillan, S.~Jolliet, A.~Bottino, P.~Angelino, T.M. Tran, and L.~Villard.
\newblock Rapid {F}ourier space solution of linear partial integro-differential
  equations in toroidal magnetic confinement geometries.
\newblock {\em Computer Physics Communications}, 181, 2010.

\bibitem{Polevoi2021}
A.R. Polevoi, A.~Loarte, R.~Bilato, N.~Gorelenkov, Ye.O. Kazakov,
  E.~Polunovskiy, A.~Tchistiakov, E.~Fable, V.~Kiptily, A.V. Krasilnikov, A.Y.
  Kuyanov, R.~Nazikian, S.D. Pinches, and M.~Schneider.
\newblock Impact of suprathermal ions on neutron yield in the pre-{DT} phase of
  {ITER} operation.
\newblock {\em Nuclear Fusion}, 61(7):076008, 2021.

\bibitem{Loarte2021}
A.~Loarte, A.R. Polevoi, M.~Schneider, S.D. Pinches, E.~Fable, E.~Militello
  Asp, Y.~Baranov, F.~Casson, G.~Corrigan, L.~Garzotti, D.~Harting, P.~Knight,
  F.~Koechl, V.~Parail, D.~Farina, L.~Figini, H.~Nordman, P.~Strand, and
  R.~Sartori.
\newblock H-mode plasmas in the pre-fusion power operation 1 phase of the
  {ITER} research plan.
\newblock {\em Nuclear Fusion}, 61(7):076012, 2021.

\bibitem{ASTRA}
G.V. Pereverzev and P.N.Yushmanov.
\newblock {ASTRA} automated system for transport analysis in a tokamak.
\newblock {\em Max-{P}lanck {IPP} Report}, 5(98), 1991.

\bibitem{Imbeaux2015}
F.~Imbeaux, S.D. Pinches, J.B. Lister, Y.~Buravand, T.~Casper, B.~Duval,
  B.~Guillerminet, M.~Hosokawa, W.~Houlberg, P.~Huynh, S.H. Kim, G.~Manduchi,
  M.~Owsiak, B.~Palak, M.~Plociennik, G.~Rouault, O.~Sauter, and P.~Strand.
\newblock Design and first applications of the {ITER} integrated modelling {\&}
  analysis suite.
\newblock {\em Nuclear Fusion}, 55(12):123006, 2015.

\bibitem{Vannini2022}
F.~Vannini, A.~Biancalani, A.~Bottino, T.~Hayward-Schneider, P.~Lauber,
  A.~Mishchenko, E.~Poli, B.~Rettino, G.~Vlad, X.~Wang, and the ASDEX
  Upgrade~team.
\newblock Gyrokinetic modelling of the {Alfv\'en} mode activity in {ASDEX
  Upgrade} with an isotropic slowing-down fast-particle distribution.
\newblock {\em Nuclear Fusion (sub.)}, 2022.

\bibitem{Rettino2022}
B.~Rettino, T.~Hayward-Schneider, A.~Biancalani, A.~Bottino, Ph. Lauber,
  I.~Chavdarovski, F.~Vannini, and F.~Jenko.
\newblock Gyrokinetic modelling of anisotropic energetic particle driven
  instabilities in tokamak plasmas.
\newblock {\em Nuclear Fusion (sub.)}, 2022.

\bibitem{CHEASE}
H.~L{\"u}tjens, A.~Bondeson, and O.~Sauter.
\newblock The {CHEASE} code for toroidal {MHD} equilibria.
\newblock {\em Computer Physics Communications}, 97, 1996.

\bibitem{Hayward-Schneider2021}
T.~Hayward-Schneider, Ph. Lauber, A.~Bottino, and Z.X. Lu.
\newblock Global linear and nonlinear gyrokinetic modelling of {Alfv\'en}
  eigenmodes in {ITER}.
\newblock {\em Nuclear Fusion}, 61:036045, 2021.

\bibitem{Lauber2009}
Ph. Lauber, M.~Br{\"u}dgam, D.~Curran, V.~Igochine, K.~Sassenberg,
  S.~G{\"u}nter, M.~Maraschek, M.~Garc{\'i}a-Mu{\~n}oz, N.~Hicks, and the ASDEX
  Upgrade~Team.
\newblock Kinetic {A}lfv\'en eigenmodes at {ASDEX U}pgrade.
\newblock {\em Plasma Physics and Controlled Fusion}, 51(12), 2009.

\bibitem{Kleiber2021}
R.~Kleiber, M.~Borchardt, A.~K{\"o}nies, and C.~Slaby.
\newblock Modern methods of signal processing applied to gyrokinetic
  simulations.
\newblock {\em Plasma Physics and Controlled Fusion}, 63(3):035017, 2021.

\bibitem{Lauber2013}
{Ph}. {Lauber}.
\newblock Super-thermal particles in hot plasmas -- kinetic models, numerical
  solution strategies, and comparison to tokamak experiments.
\newblock {\em Physics Reports}, 533(2), 2013.

\bibitem{Van_Zeeland2016}
M.A.~Van Zeeland, W.W. Heidbrink, S.E. Sharapov, D.~Spong, A.~Cappa, Xi~Chen,
  C.~Collins, M.~Garc{\'{\i}}a-Mu{\~{n}}oz, N.N. Gorelenkov, G.J. Kramer,
  P.~Lauber, Z.~Lin, and C.~Petty.
\newblock Electron cyclotron heating can drastically alter reversed shear
  alfv{\'e}n eigenmode activity in {DIII}-d through finite pressure effects.
\newblock {\em Nuclear Fusion}, 56(11):112007, 2016.

\bibitem{Heidbrink2008}
W.~W. Heidbrink.
\newblock Basic physics of {\Alfv} instabilities driven by energetic particles
  in toroidally confined plasmas.
\newblock {\em Physics of Plasmas}, 15(5), 2008.

\bibitem{Zonca1996}
Fulvio Zonca, Liu Chen, and Robert~A Santoro.
\newblock Kinetic theory of low-frequency alfv{\'{e}}n modes in tokamaks.
\newblock {\em Plasma Physics and Controlled Fusion}, 38(11), 1996.

\bibitem{Zonca1998}
Fulvio Zonca, Liu Chen, Robert~A Santoro, and J~Q Dong.
\newblock Existence of discrete modes in an unstable shear alfv{\'e}n
  continuous spectrum.
\newblock {\em Plasma Physics and Controlled Fusion}, 40(12), 1998.

\bibitem{Popa_MSc}
V.-A. Popa.
\newblock Workflow-based energetic particle stability analysis of projected
  {ITER} plasmas.
\newblock M.{S}c. thesis, Technische Universit\"at M\"unchen, 2021.

\bibitem{HELENA}
G.T.A. Huysmans, J.P. Goedbloed, and W.~Kerner.
\newblock Isoparametric bicubic {H}ermite elements for solution of the
  {G}rad-{S}hafranov equation.
\newblock {\em Proc. CP90 Conf. on Comp. Phys. Proc.}, page 371, 1991.

\end{thebibliography}
